\documentclass[twocolumn, tighten, dvipsnames]{aastex701}

\received{2025 August 23}
\revised{2025 September 18}
\accepted{2025 October 3}
\submitjournal{AJ}

\shorttitle{Methanol in TW Hya}
\shortauthors{Ilee et al.}

\usepackage{xspace}
\usepackage[version=4]{mhchem}

\def\arraystretch{1.3}

\begin{document}

\title{Methanol emission tracing ice chemistry and dust evolution in the TW Hya protoplanetary disk}

\author[0000-0003-1008-1142]{John~D.~Ilee} 
\affiliation{School of Physics and Astronomy, University of Leeds, Leeds, UK, LS2 9JT}
\email[show]{J.D.Ilee@leeds.ac.uk}  

\author[0000-0001-6078-786X]{Catherine~Walsh}
\affiliation{School of Physics and Astronomy, University of Leeds, Leeds, UK, LS2 9JT}
\email[show]{C.Walsh1@leeds.ac.uk}  

\author[0000-0002-0150-0125]{Jenny~K.~Calahan}
\affiliation{Center for Astrophysics \textbar\ Harvard \& Smithsonian, 60 Garden St., Cambridge, MA 02138, USA}
\email{jenny.calahan@cfa.harvard.edu}

\begin{abstract}
Methanol (\ce{CH3OH}) ice is abundant in space and is a key feedstock for seeding chemical complexity in interstellar and circumstellar environments. 
Despite its ubiquity, gas-phase methanol has only been detected in one disk around a Solar-type star to date, TW Hya.  
Here we present new high sensitivity ($\sim 1$~mJy/beam) observations of TW Hya with ALMA that detect four individual transitions of gas-phase methanol spanning upper level energies from 17 to 38~K. 
We confirm the presence of gas-phase methanol in the luke-warm molecular layer of the disk ($35.9^{+25.9}_{-10.6}$~K) and with a disk-integrated column density of $1.8^{+1.3}_{-0.5}\times 10^{12}$~cm$^{-2}$. 
A radially-resolved analysis suggests that the gas-phase methanol is centrally compact, peaking within the spatial extent of the mm-sized dust grains ($\lesssim 80$~au). 
Static gas-grain chemical disk models confirm photodesorption as an important mechanism releasing methanol into the gas phase, with the column density further boosted by the inclusion of grain-surface chemistry, reactive desorption, and an increase in dust-grain surface area assuming fractal grains. 
However, no model can fully reproduce the observed column density nor the radial distribution, and we suggest that the inclusion of dynamic processes such as vertical mixing and radial drift would be required to do so.  
Our results demonstrate that the abundance and distribution of the precursors for complex chemistry in the planet-forming regions around Solar-type stars is ultimately controlled by the interplay of grain surface chemistry coupled with the evolution of dust in their disks.
\end{abstract}


\keywords{Protoplanetary disks(1300) ---
Astrochemistry(75) ---
Interstellar molecules(849) ---
Planet formation(1241)}

\section{Introduction} 
\label{sec:intro}

Methanol (\ce{CH3OH}) is the fourth most abundant ice species observed 
in interstellar clouds \citep[e.g.,][]{Boogert2015} and is considered to be an important 
feed-stock for growing molecular complexity in space, including as a building block for potentially prebiotic molecules such as amino acids and sugars \citep[e.g.,][]{Allamandola1988, Bennett2007, Oberg2009, Modica2010, Chen2013, Bergantini2018, Zhu2020}.
Gas-phase methanol is almost ubiquitously observed across the diverse 
range of environments in interstellar and circumstellar space; within cold, starless cores \citep[e.g., see a recent survey by][]{Spezzano2022}, in the warm cocoons around low-mass \citep[e.g.,][]{Maret2005} and high-mass protostars \citep[e.g.,][]{vandertak2000}, in the shocks emanating from the impact of jets from young stars on their surroundings \citep[e.g.,][]{Bachiller1995}, and even in the interstellar media of other galaxies \citep[e.g.,][]{Henkel1987}.  
Methanol masers have long been used as indicators of the onset of star-formation activity within massive, dense, molecular cloud complexes where other observations cannot penetrate \citep[e.g.,][]{Breen2010}.
Notably, methanol ice is also a major component of cometary ices, possessing an abundance of 0.6\% to 6.2\% relative to water ice, revealing an intriguing chemical link between interstellar and cometary ices \citep[see, e.g., the review by][]{BockeleeMorvan2017}.  
Indeed, chemodynamical models have suggested that young disks may inherit a substantial fraction of interstellar methanol indicating that disks already begin their chemical evolution with this important feed-stock for seeding further complexity \citep[][]{Drozdovskaya2014}.

Gas-phase methanol has now been detected in five (Class II) protoplanetary disks to date: 
TW Hya \citep{Walsh2016}, HD~100546 \citep{Booth2021,Booth2024HD100546,Evans2025},  IRS~48 \citep{vanderMarel2021,Booth2024irs48}, HD~164192 \citep{Booth2023}, and HD~100453 \citep{Booth2025}.  
There are also non-detections and upper limits determined for the disks around HD~163296 \citep{Carney2019} and MWC~480 \citep{Yamato2024}.  
All of these disks, with the exception of TW Hya, are disks around Herbig Ae/Be stars.
These detections were only possible because of the increase in line sensitivity provided by the Atacama Large Millimeter/submillimeter Array (ALMA) compared with previous facilities. 
The challenges related to detecting gas-phase methanol (and indeed other complex organic molecules, henceforth referred to as COMs) are 
due to the physical conditions present in protoplanetary disks, as well as their small apparent size (typically a few arcseconds at most) relative to 
other astrophysical environments (e.g., clouds, or protostellar envelopes).  
The vast majority of the dust and gas in most protoplanetary disks is cold ($\ll 100$~K) and so molecules 
such as \ce{H2O} and \ce{CH3OH} are frozen as ice on dust grains beyond the snowline, which 
lies at around 1 -- 2~au for disks around T Tauri stars and between 5 -- 15 au for disks around Herbig~Ae/Be stars \citep[e.g.,][]{Agundez2018}, 
with some variation in these numbers expected due to the presence of disk sub-structures (e.g., cavities).

The successful detections of gas-phase methanol in the inner regions of now four disks around Herbig~Ae/Be stars is likely due to their possession of dust cavities. This allows direct irradiation of the inner cavity edge, exposing the ices in the midplane to heat, thereby driving efficient thermal desorption and detectable gas masses of methanol \citep[see, e.g., the discussion in][]{Booth2025}. 
This picture is consistent with the non-detections of gas-phase methanol in the Herbig~Ae/Be disks of HD~163296 and MWC~480 which do not possess large dust cavities, although the innermost region of MWC~480 does show emission from other large O-bearing molecules, e.g.\ dimethyl ether \citep{Yamato2024}.
Nevertheless, TW Hya remains the only disk around a Solar-type star in which gas-phase methanol has been robustly detected to date. 

The origin of gas-phase methanol in TW Hya, differs from that from the Herbig Ae/Be disks discussed above due to its snowline lying at only a few au from the central star. For gas-phase methanol to exist in the gas-phase beyond the snow line, there must be a non-thermal desorption mechanism at work that is able to release the molecule from the ice at a sufficient rate to counteract (re)freeze-out onto dust grains. An alternative possibility is that there is sufficient dust depletion in the disk molecular layers that the freeze-out timescales become longer that the disk life-time, effectively switching off grain-surface interactions and chemistry. In this latter case, gas-phase methanol would need to have an efficient gas-phase formation mechanism to counteract destruction in the molecular layer via UV and X-ray photons, and ion-molecule chemistry. 

Due to its proximity \citep[60~pc;][]{Gaia2023DR3} and well characterised physical structure, TW Hya provides a unique laboratory to study the non-thermal desorption mechanism of methanol (and other COMs) in a protoplanetary disk around a Sun-like star, and in particular to explore the relative importance of photodesorption, X-ray desorption, and reactive desorption in setting the abundance of gas-phase methanol in the disk molecular layer. 
Determining the dominant chemical origin will then help to better constrain the abundance and distribution of the methanol ice reservoir to compare with the cometary record in the Solar System.

Here, we present new ALMA observations of rotational lines of gas-phase methanol in TW Hya that span a range of excitation energies from 16 to 68~K.  
These data enable us, for the first time, to constrain the excitation temperature and column density of methanol in the TW Hya disk and to perform a spatially resolved analysis to constrain the radial distribution.  
For the previous detection reported in \citet{Walsh2016}, whilst several lines were targeted, the lines needed to be stacked in order to successfully image the data and to confirm detection (although subsequent analysis confirmed the detection of the individual lines in $uv$ space via matched filtering, see \citealt{Loomis2018b}). 
We compare the empirically derived column density and excitation temperature with the results of a full chemical model in which we test the impact of varying the rates of non-thermal desorption on the resulting abundance and distribution of gas-phase methanol, and explore the possibility of an active gas-phase route in the case of dust depletion in the molecular layer.

In Section~\ref{sec:obs} we describe our observations, in Section~\ref{sec:results} we present the results of our imaging analysis including radial profiles, in Section~\ref{sec:discussion} we compare our data driven results with those from a full chemical model, in Section~\ref{sec:discussion2} we discuss the possible chemical origins of gas-phase methanol in TW Hya, and in Section~\ref{sec:summary} we summarise our work and state our conclusions.

\section{Observations}
\label{sec:obs}

\begin{figure*}
    \centering
    \includegraphics[width=\textwidth,trim={0 0.5cm 0 2cm}, clip]{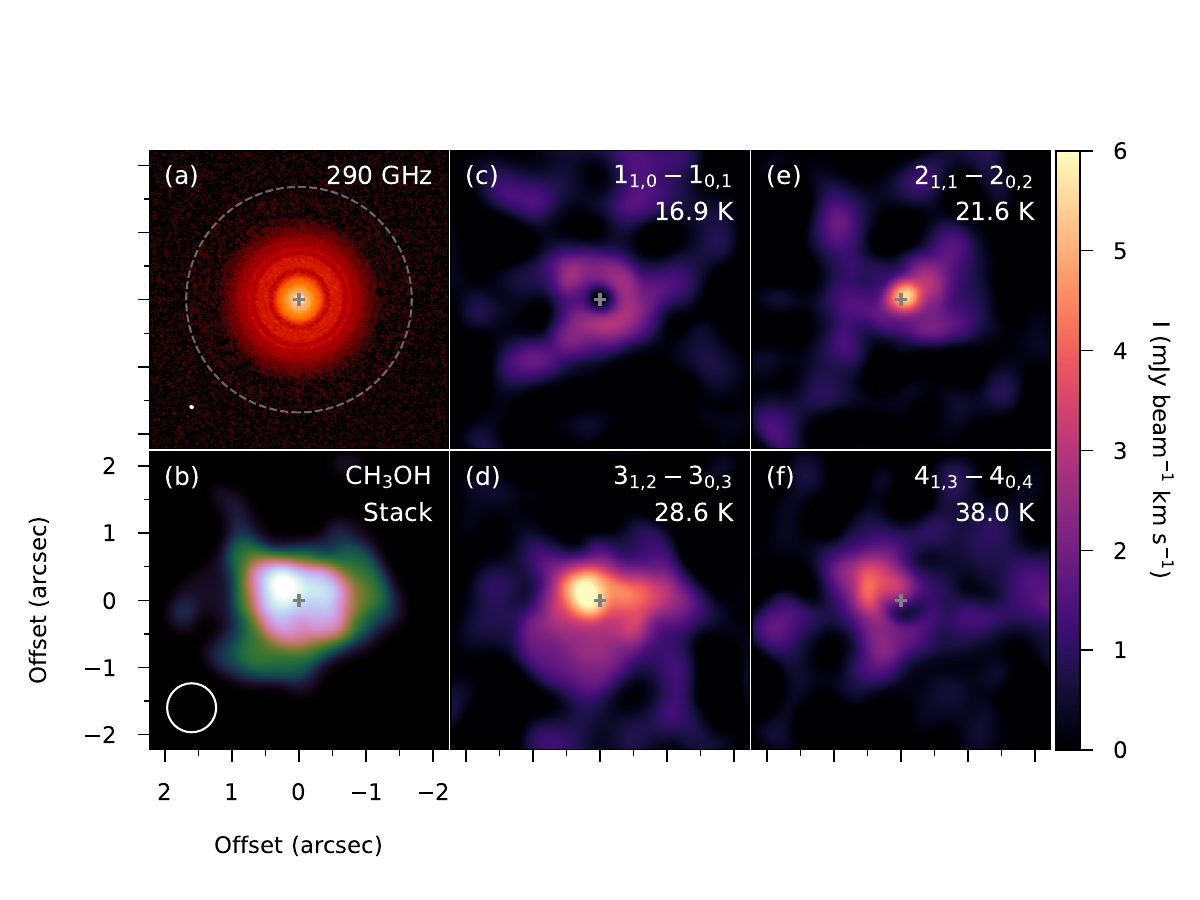}
    \caption{Morphology of the methanol (\ce{CH3OH}) emission toward TW~Hya. \textbf{(a)} 290\,GHz continuum emission from \citet{Huang2018} overlaid with the dust disk extent derived from our observations in \citet{Ilee2022} (dashed ellipse). \textbf{(b)} Integrated intensity map of the $0\farcs7$ stacked methanol transitions. \textbf{(c-f)} As (b), but for the individually detected transitions.}
    \label{fig:moments}
\end{figure*}

TW Hya was observed by ALMA in Band 7 on the 3rd of December 2016 for 5.7 hours in configuration C40-4 under project code 2016.1.00464.S (P.I.~C.~Walsh) with a precipitable water vapour measurement of 1.0~mm.  Baseline lengths ranged from 15--704~m with 39--46 antennas depending on execution.  Quasar J1037$-$2934 was used as both a phase and flux calibrator, while J1058$+$0133 was used as a bandpass calibrator. The correlator was configured for dual polarisation across one continuum and eight line-containing spectral windows across a frequency range of 291--307~GHz with spectral resolutions of 32.0 and 0.13~km~s$^{-1}$, respectively.  Table \ref{tab:molecular} details the targeted methanol transitions.

Data (self-)calibration and imaging were performed with CASA version 5.6.1 \citep{CASA}.  A continuum visibility measurement set was created by combining all channels in the continuum spectral window after flagging those containing strong line emission.  Two rounds of phase self-calibration and one round of amplitude self-calibration were undertaken, improving the peak signal-to-noise ratio in the continuum image by a factor of 14.  The phase centre was set to $11^{\rm h}01^{\rm m}51\fs811$,  $-034\degr42\arcmin17\farcs267$ throughout.  The continuum self-calibration solutions were applied to the measurement sets containing line emission.  An analysis of the continuum emission is presented in \citet{Ilee2022}. 

\subsection{Matched filtering}

Due to the weak nature of the targeted methanol transitions, we initially processed the measurement sets of the line-containing spectral windows using a matched filter technique, which is fully described in \citet{Loomis2018b}.  Briefly, this utilises a known morphological and kinematic structure of a source in order search for line emission directly in the Fourier domain, resulting in an improvement of signal-to-noise ratio when compared with imaging alone. We experimented with a variety of Keplerian filters with varying radial extents from 50 -- 200~au, finding that the strongest filter responses for the methanol transitions came from filters of radius 50~au, demonstrating the compact nature of the emission.  Appendix \ref{sec:filter} shows the results of the matched filter for all spectral windows in the observations. 

\subsection{Imaging}

\begin{figure*}
    \centering
    \includegraphics[width=\textwidth, trim={0 1.75cm 0 0cm}, clip]{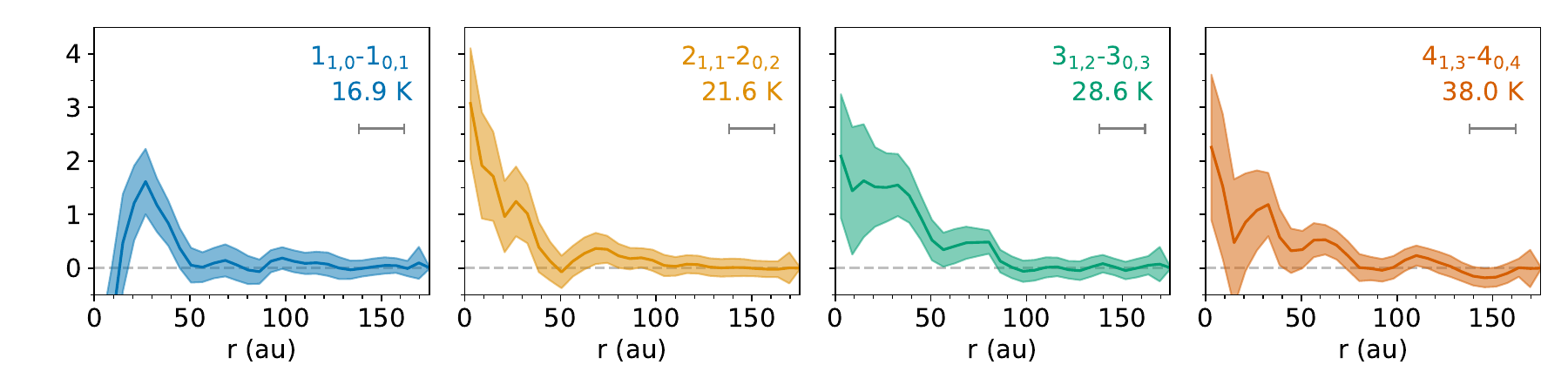}
    \includegraphics[width=\textwidth]{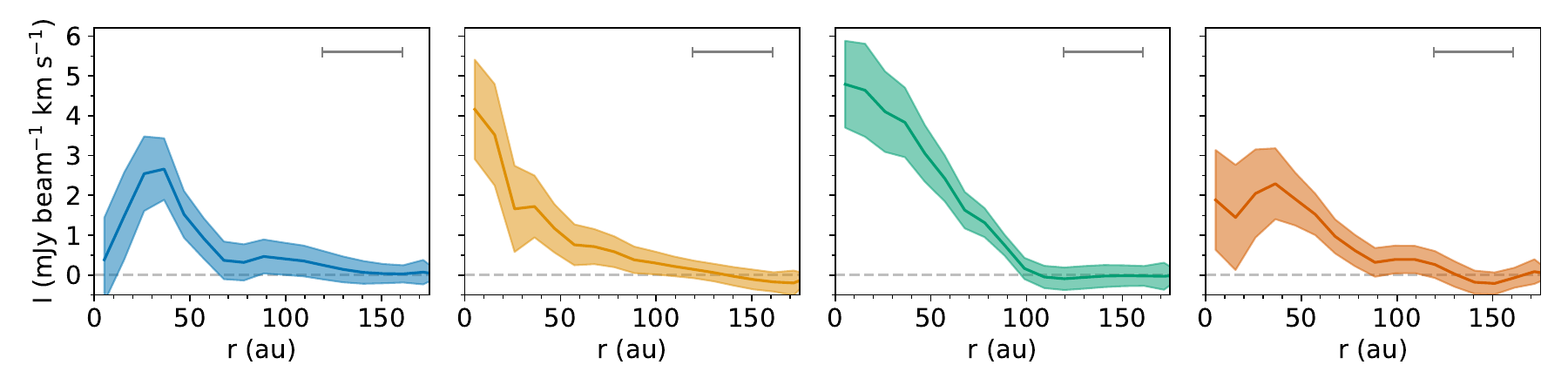}
    \caption{Radial profiles of emission for the detected methanol (\ce{CH3OH}) transitions.  The upper panels show the 0\farcs4 observations, while the lower panels show the 0\farcs7 observations.  Transitions are labelled with $E_\mathrm{u}$, shaded regions indicate 1$\sigma$ uncertainties} and the beam FWHM is shown with a horizontal bar.
    \label{fig:radial_profile}
\end{figure*}

We then imaged the detected transitions with the CASA \texttt{tclean} task, using Keplerian masks calculated from the known geometry of the TW Hya disk (an inclination of 5\degr~and position angle of 152\degr; \citealt{Huang2018}) to select regions in position-position-velocity space expected to have emission.  After experimenting with a range of Briggs robust parameters ($R$) and different levels of $uv$-tapering, we produced two sets of images for subsequent analysis.  Our high resolution case, used to investigate radially-resolved emission while still allowing adequate detections of all transitions, was imaged with $R=0.0$ and no taper to achieve a $0.4\arcsec$ beam major axis (rms per channel = 1.15~mJy\,beam$^{-1}$).  Our high sensitivity case, used to investigate extended disk-integrated emission, was imaged with $R=0.5$ and a $uv$-taper to achieve a $0.7\arcsec$ beam major axis (rms per channel = 1.45~mJy\,beam$^{-1}$).  In both cases, the beams were subsequently circularised using the \texttt{imsmooth} task (although we note that the original size of the minor axes were within 10\% of the major axes due to the favourable declination of TW~Hya during the observations). We adopted a common velocity resolution of 0.2\,km\,s$^{-1}$ in order to improve signal-to-noise compared to the native spectral resolution.  Channel maps for both sets of images are presented in Appendix \ref{sec:channelmaps}.

\subsection{Results}
\label{sec:results}

The matched filter results (see Appendix \ref{sec:filter}) confirm the successful detections ($\gtrsim 4\sigma$) of the rotational lines of gas-phase methanol spanning upper level energies from 17 to 38~ K. 
We do not detect the higher energy line with an upper level energy of 64~K (see Table~\ref{tab:molecular}) which already hints that the rotational temperature of gas-phase methanol in TW Hya is low.
In this section, we present data products derived from the methanol image cubes of the detected lines, including integrated intensity maps, disk-integrated flux measurements, radial emission profiles, and rotational diagram analysis using the relative strengths of the detected transitions.  

\begin{deluxetable*}{lcccccc}
\tablecaption{List of targeted methanol (\ce{CH3OH}) transitions and molecular spectroscopic data\tablenotemark{$\dag$} assumed during our calculations.  Also given are measurements of the peak matched filter response ($\sigma_{\mathrm{f}}$) and disk-integrated fluxes ($S_{\nu} \Delta v$) obtained from the $0\farcs7$ observations. \label{tab:molecular}}
\tablewidth{0pt}
\tablehead{\colhead{Transition} & \colhead{Frequency} & \colhead{$E_\mathrm{u}$} & \colhead{log$_{10}(A_{\rm ij})$} & \colhead{$g_\mathrm{u}$} & \colhead{$\sigma_{\mathrm{f}}$} & \colhead{$S_{\nu} \Delta v$} \\
  &  \colhead{(GHz)} & \colhead{(K)} & & & & \colhead{(mJy\,km\,s$^{-1}$)} }
\startdata
1(1,0)-1(0,1) A  &    303.3669  &  16.9 & $-3.49308$  & 12 & 5.8 & $14.3 \pm 2.5$ \\
2(1,1)-2(0,2) A  &    304.2083  &  21.6 & $-3.49028$  & 20 & 7.8 & $12.1 \pm 2.1$ \\
3(1,2)-3(0,3) A  &    305.4735  &  28.7 & $-3.48628$  & 28 & 7.6 & $24.6 \pm 2.7$ \\
4(1,3)-4(0,4) A  &    307.1659  &  38.0 & $-3.48073$  & 36 & 4.9 & $13.0 \pm 2.7$ \\
6(1,5)-5(1,4) A  &    292.67289  &  63.7 & $-3.97483$  & 52 & $<3$ & $<7.4$ \\
\enddata
\tablenotetext{$\dag$}{obtained from the Cologne Database for Molecular Spectroscopy (CDMS,  \url{https://cdms.astro.uni-koeln.de/cdms/portal/})}
\end{deluxetable*}

\subsubsection{Integrated intensity maps}

To investigate the general morphology of the methanol emission, we generated integrated intensity (zeroth moment) maps for each transition with the \texttt{bettermoments} package \citep{teague_bettermoments} using Keplerian masks.  Such an approach has the benefit of significantly improving the fidelity of intensity maps generated from weaker transitions (see \citealt{Law2021rad}, their Appendix A, for a full discussion of this approach).  Figure \ref{fig:moments} shows the integrated intensity maps of the methanol transitions generated from the 0\farcs7 images compared to high-resolution dust continuum observations of TW~Hya at a similar frequency \citep[290~GHz; from][]{Huang2018}.

The individual transitions appear to show a range of different morphologies including arcs/rings, central peaks and more extended emission.  The $1_{1{,}0}$--$1_{0{,}1}$ transition appears to emit from a ring at approximately 30\,au, while the $2_{1{,}1}$--$2_{0{,}2}$ appears anti-correlated to this with a central peak.  The $3_{1{,}2}$--$3_{0{,}3}$ transition, which is strongest, is also centrally peaked but shows an extended shelf of emission out to approximately 60\,au.  The $4_{1{,}3}$--$4_{0{,}4}$ transition appears to be intermediate between these two cases, with a central depression surrounding by an asymmetric arc to the North-East.  Figure \ref{fig:moments}b shows the integrated intensity map of these detected transitions stacked in the image plane.  From this, it is clear that, in general, the methanol emission is entirely contained within the extent of the millimetre dust disk (out to between 60--90\,au).

\subsubsection{Radial profiles}

To further investigate the emission morphology, we created radial emission profiles for each transition using the \texttt{GoFish} package \citet{teague_gofish}, with a grid set to the inclination (5\degr) and position angle (152.0\degr) of TW~Hya \citep{Huang2018}.  We assumed a flat emission layer (e.g., $z/r = 0$), but note that due to the near face-on orientation of the disk, other choices do not affect the resulting emission profiles.  Figure \ref{fig:radial_profile} shows the profiles derived for both the high resolution 0\farcs4 and high sensitivity 0\farcs7 image cubes.   

The profiles confirm the slightly different morphologies of emission for the different transitions.  A central depression is clearly seen in the $1_{1{,}0}$--$1_{0{,}1}$ transition, while the intensity of the other transitions decreases monotonically from either a central peak or plateau.  The profiles also confirm the compact nature of the emission, with only significant signal detected out to between 50--100\,au depending on transition.  There may be hints of radial substructure seen in the 0\farcs4 profiles between approximately 50--70\,au, however the size of these features is comparable to the beamsize, and their significance is relatively low.  

\subsubsection{Rotational diagrams}

The detection of multiple methanol transitions that span a range of upper state energy levels ($E_{\rm u} = 16.8$--38.0\,K) allows us to use their relative strengths to quantify the physical conditions of the gas in the emitting region.  To this end, we performed a rotational diagram analysis.  In the first instance, we measure disk integrated fluxes using the line shifting and stacking technique provided by the \texttt{integrated\_spectrum()} function within \texttt{GoFish} \citep{teague_gofish}.  This process deprojects the rotation profile of the disk and combines Doppler-shifted emission onto a common centroid velocity reference frame, yielding a single disk-integrated spectrum from which to measure the strength of each transition.  In the case of weak lines such as those presented here, these spectra provide an increase in the signal-to-noise ratio when compared with other techniques.  The resulting disk-integrated line flux measurements are given in Table \ref{tab:molecular}.

Our calculation of a rotational diagram from these measurements is fully outlined in \citet{Ilee2021}.  Briefly, this involves determining a rotational temperature, $T_\mathrm{rot}$, and total column density, $N_T$, from a linear least squares regression deriving the best fitting slope and intercept of the relationship between the relative strength of transitions as a function of their upper state energy level, $E_{\rm u}$. This assumes that local thermodynamic equilibrium (LTE) governs the level populations, which is the case for the strong lines of gas-phase methanol targetted in this work \citep[see the analysis conducted in][]{Parfenov2017}. We note that methanol exists in two nuclear spin states, A-type (all proton spins are parallel) and E-type (one proton spin is antiparallel).  While all of our targeted transitions correspond to A-type methanol, the molecular spectroscopic data we use intrinsically accounts for these multiple states, meaning that all column densities we report reflect the total (A-type plus E-type) abundance (see, e.g., \citealt{Lees1973} and \citealt{Muller2021}).  For the calculation of the column density, we assume an emitting area based on a region 80\,au in radius since this encompasses all of the significant emission from each of the targeted methanol transitions.

\begin{figure}
    \centering
    \includegraphics[width=\columnwidth]{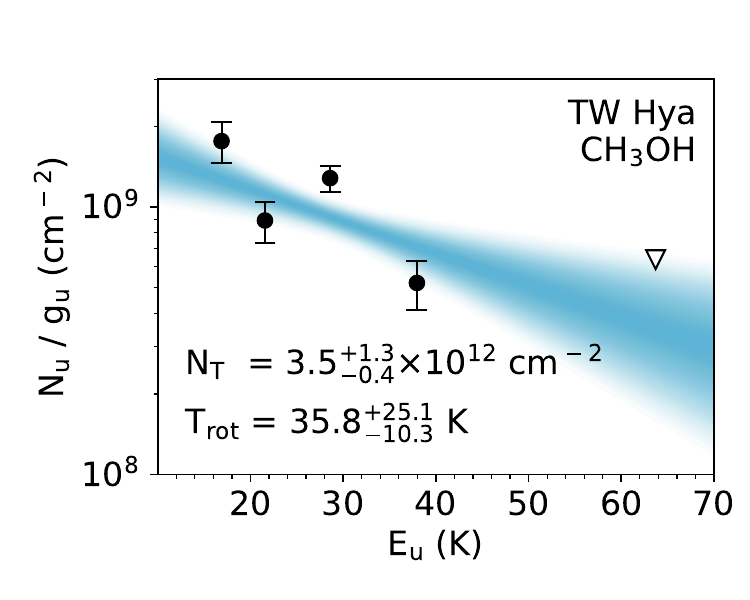}
    \caption{Rotational diagram calculated from the disk-integrated values of the 0$\farcs$7 observations.  Upper limits are denoted with a downward triangle, random draws from the corresponding posterior probability distribution are shown in blue, and median values with their uncertainties are labelled.}
    \label{fig:rotational}
\end{figure}

Figure \ref{fig:rotational} shows the rotational diagram calculated from the disk-integrated line flux measurements from the 0\farcs7 images.  The diagram is relatively well fit by a single component, with median values of column density and rotational temperature from the posterior probability distribution of the MCMC fitting procedure are $N_{\rm T} = 1.8^{+1.3}_{-0.5}\times10^{12}$~cm$^{-2}$ and $T_{\rm rot} = 35.9^{+25.9}_{-10.6}$\,K, respectively (where uncertainties correspond to the 16$^{\rm th}$--84$^{\rm th}$ percentile of the distribution).  All transitions have calculated optical depth values of $\tau < 10^{-2}$ demonstrating they are optically thin. 

\begin{figure}
    \centering
    \includegraphics[width=\columnwidth]{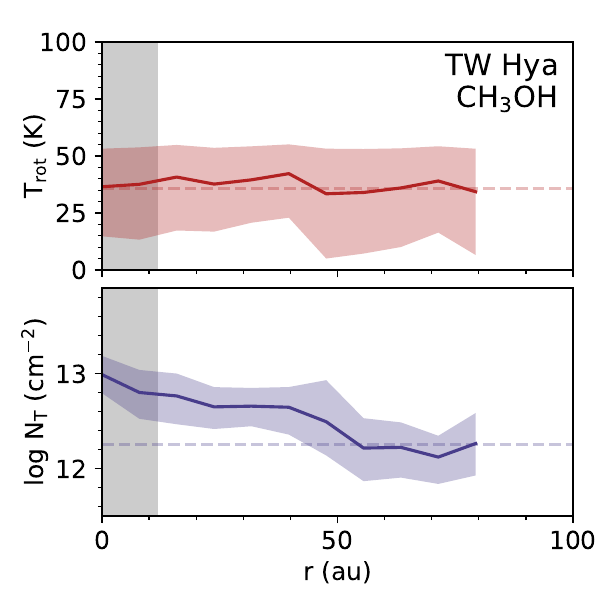}
    \caption{Radially resolved rotational diagram for the 0$\farcs$4 observations, where shaded regions indicate uncertainties obtained from the 16$^{\rm th}$--84$^{\rm th}$ percentile of the posterior distribution. Dashed lines show values derived from the disk integrated analysis (see Figure~\ref{fig:rotational}). The vertical grey region indicates a radial extent of half the beam size.}
    \label{fig:rotational_radial}
\end{figure}

We can extend this analysis to investigate the radial behaviour of column density and excitation temperature using the radial profiles of the higher resolution 0\farcs4 images, and calculating a rotational diagram for each radial bin.  Figure \ref{fig:rotational_radial} shows the corresponding $N_{\rm{T}}(r)$ and $T_{\mathrm{rot}}(r)$ profiles derived from this analysis.  The rotational temperature shows little radial dependence out to 80\,au (beyond which there is insignificant signal from multiple transitions), and agrees very well with the disk-integrated value.  The lack of rotational temperature variation within 30\,au (despite a drop in intensity of the lowest energy $1_{1{,}0}$--$1_{0{,}1}$ transition) can be explained by the fact that the gradient of the fit in these regions is driven primarily by the other strongly detected higher energy transitions. The column density profile monotonically decreases from a central peak of $10^{13}$~cm$^{-2}$ and reaches agreement with the disk-integrated value beyond a radius of approximately 50\,au.  There appears to be little in the way of substructure in either profile, with any variations well within the associated uncertainties for each quantity.

\section{Gas-grain chemical models of the disk of TW~Hya}
\label{sec:discussion}

\subsection{The chemical origin of methanol in TW Hya}

Both our disk-integrated and radially resolved rotational diagram analysis indicates that the methanol emission in TW~Hya is rotationally cold, with $T_{\rm rot} \approx~36$~K across the full radial extent of the emission (out to $\sim$80~au).  We can compare this to the evaporation temperature expected if the methanol were being thermally desorbed from icy dust grains in the disk.  Depending on assumptions regarding the composition of the icy grain surface, \citet{Minissale2022} find evaporation temperatures for methanol between 130--170\,K.  Such values are significantly higher than our derived rotational temperatures, indicating that the origin of the gas phase methanol in TW~Hya cannot be due to direct thermal desorption.  Instead, it must originate from either non-thermal desorption processes or form in the gas phase directly.

Methanol ice is efficiently formed from sequential hydrogenation of CO ice at low temperatures (see, e.g.~\citealt{Hiraoka1994}, \citealt{Watanabe2002}, and \citealt{Fuchs2009}). 
On the other hand, it has been proposed for some time that gas-phase formation of methanol is inefficient because it relies on the formation of a protonated complex formed by radiative association (which is typically slow), 
\begin{align}
\ce{CH3+ + H2O &-> CH3OH2+ + h\nu}, \nonumber
\end{align}
that then undergoes dissociative recombination with an electron which results in fragmentation of the complex \citep[][]{Garrod2006,Geppert2006},
\begin{align}
\ce{ &-> CH2 + H2O + H} \nonumber \\
\ce{  &-> CH3 + H2O }\nonumber \\
\ce{CH3OH2+ + e-  &-> CH3 + OH + H }\\
\ce{  &->  H2CO + H2 + H }\nonumber\\
\ce{  &-> CH3OH + H }\nonumber
\end{align}
with only 3\% of dissociations leading to intact methanol. A potential solution to this issue is that reactions of the protonated complex with a proton acceptor such as \ce{NH3} can provide a route to intact gas-phase methanol \citep{Rodgers2001}; however, one would need a sufficient abundance of \ce{NH3} available in the gas in the molecular layer, which brings us back to the requirement for the release of molecules with high binding energies from an ice reservoir via non-thermal desorption.

Non-thermal desorption from ices can arise in disks via photodesorption, where photons can be produced internally in media due to the interaction of cosmic rays and /or X-rays with molecular hydrogen or originate externally from, e.g., the host star.  
A second non-thermal desorption mechanism is reactive desorption, where a proportion of the excess energy created in bond formation goes into releasing the product of a grain surface association reaction into the gas-phase (see the reviews by \citealt{Oberg2016} and \citealt{Cuppen2017}). 
There are also alternative explanations presented in the literature such as grain-grain collisions \citep[e.g.,][]{Kalvans2022} and rotational desorption due to radiative torques \citep[e.g.,][]{Hoang2019}.  
However these mechanisms are both unlikely to be occurring in the outer regions of disks where the grain relative velocities are low, and where the radiation field is too weak, respectively.  
It has also been proposed that \ce{CH3OH} may be able to co-desorb with CO ice at and around the location of the CO snowline and surface; however, sensitive laboratory experiments could only put an upper limit on the process of $\approx 10^{-6}$ methanol molecules per CO molecule \citep{Ligterink2018}.  

Similar to the issue with gas-phase formation of methanol, experiments have shown that pure and mixed methanol ices fragment upon irradiation with UV photons.  
For UV irradiated ices, a desorption yield of $10^{-5} - 10^{-6}$ molecules per photons has been measured \citep[][]{Bertin2016,Cruzdiaz2016}, which can be compared with the higher photodesorpion yields for more simple species such as CO, \ce{N2}, \ce{CO2}, and \ce{H2O} ($10^{-4} - 10^{-2}$ molecules per photon; see reviews by \citealt{Oberg2016} and \citealt{Cuppen2017}).
On the other hand, for X-ray irradiated ices, intact methanol desorption has been detected with a much higher yield of $10^{-2}$ molecules per 564 eV X-ray photon \citep[][]{Basalgete2021}.
The simulation of such processes in astrochemical models is complicated by the fact that irradiation induces a cascade of processes on the molecular level that includes excitation, ionisation, dissociation, diffusion, recombination, reaction, and desorption.

\subsection{Description of the chemical model}

We next use a gas-grain chemical model combined with an observationally-constrained physical structure for the disk around TW Hya \citep[from][]{calahan21}, to compute the abundance of gas-phase and ice-phase methanol for a suite of models in which we include/exclude various chemical ingredients. 
We do this to investigate which chemical processes are most important for setting the abundance and distribution of gas-phase methanol in the disk of TW~Hya.

As already discussed, the most likely explanation is that the gas-phase methanol detected in TW~Hya arises due to non-thermal desorption from the ice mantle.  
To test this theory, and to quantify the impact of the possible chemical origins (e.g., photodesorption, X-ray desorption, and reactive desorption) on the abundance of gas-phase methanol, we run a suite of gas-grain chemical models. 
The model and network is identical to that used in \citet{walsh2015} with some additional modifications introduced to maximise the calculated abundance of gas-phase methanol, and to enable a more realistic treatment of dust properties. 
Early gas-grain models often assume a fixed dust grain size and density based on that constrained for the interstellar medium.  
However, it is now known from observations that the dust in protoplanetary disks has undergone substantial evolution through growth, settling, radial drift, and trapping \citep[see the recent review by][]{Birnstiel2024}, that will modify the local dust size and density distribution. 
Several protoplanetary disks thermochemical codes now account for these effects by adopting two dust populations, one population of ``small'' dust grains (up to 1~$\mu$m in size) and a second population of ``large'' dust grains (up to 1~mm in size), which have different total masses and scale heights, with the ``large'' grains more settled (i.e., with a smaller scale height) than the small \citep[following the parametrised approach first suggested by][]{DAlessio2006}.  
For simplicity, both dust populations are often assumed to follow the typical MRN size distribution \citep{Mathis1977}, $dn/da \propto a^{-3.5}$, where $n$ is the number of grains, and $a$ is the grain size.

The gas-grain model from \citet{walsh2015} has now been modified to ingest a dust density and size distribution which allows the available dust grain surface area per unit volume to vary through the disk which will affect the rates for accretion, desorption, and diffusion of species on the dust grain surfaces. 
Further, the code also has two additional flags to allow, i) an increase in available dust-grain surface area due to the proposed fractal nature of dust \citep[as is suggested to be the case in IM~Lup by][]{Tazaki2023}, and ii) an increase in the {\em ice-to-gas} mass ratio scaled by the {\em dust-to-gas} mass ratio, which makes the simple assumption that the ice reservoir has followed the large grains, and has become more radially and vertically concentrated in the midplane of the disk. 
The first flag scales the available dust-grain surface area by the area-to-mass ratio, $A/m$, constrained by \citet{Tazaki2023} for the disk of IM~Lup, where $A/m \simeq 1.5 \times 10^{4}~\rm{cm}^2~\rm{g}^{-1}$. 
The second flag simply scales the initial abundance of all ice species by the ratio of the input dust-to-gas mass ratio to the canonical value of 0.01, i.e., if the dust-to-gas mass ratio is reduced by a factor of 10 relative to the canonical value, then the initial abundances of ice species in that part of the disk are also reduced by a factor of 10 (and vice versa). 
The code has also been modified to treat X-ray desorption as a separate process to photodesorption, using the depth-dependent desorption yields recently measured in the laboratory for CO, \ce{H2O}, \ce{CH3OH}, and \ce{CH3CN} \citep[][]{Dupuy2018,Dupuy2021,Basalgete2021,Basalgete2023}. 
For species without an available measurement we used the yields for \ce{H2O} by default. 
Fragmentation upon photodesorption of methanol ice is already included in the code \citep[][]{Bertin2016}.

We adopt the physical structure from \citet{calahan21} for the disk of TW~Hya which was calculated using the thermochemical code, rac-2d \citep{Du2017code}. 
This model reproduces well many observations of TW Hya including the spectral energy distribution (SED), the flux in the HD $J=1-0$ line, and the radial distribution of the integrated intensity of several CO (including $^{13}$CO and C$^{18}$O) rotational lines (from $J=2-1$ to $J=6-5$). 
We present and discuss the physical structure in the next subsections. 

Our initial chemical abundances are generated from a single-point gas-grain chemical model of a cold, dark molecular cloud with a temperature of 10~K, a gas number density of $10^4$~cm$^{-3}$, a visual extinction of 10~magnitudes, and a cosmic-ray ionisation rate of $1.3 \times 10^{-17}$~s$^{-1}$, with the abundances extracted at a time of 1~Myr. 
Hence, we begin our calculations with already formed ice mantles, with fractional abundances (relative to total H nuclei) for \ce{H2O}, \ce{CO2}, \ce{CO}, and \ce{CH3OH} of $1.8 \times 10^{-4}$, $2.8 \times 10^{-6}$, $2.2 \times 10^{-5}$, and $2.3 \times 10^{-5}$, respectively, which are only adjusted through the disk when the flag for ice mass scaling is switched on. 

\subsection{Abundance of methanol gas and ice at 30~au}

To explore how the various chemical processes affect the column density of gas-phase methanol calculated for TW Hya, we first run a suite of models with different chemical ingredients, across a vertical slice of the disk at a radius of 30~au.  
We choose 30~au as this roughly coincides with the size of the beam in our observations (see Section~\ref{sec:obs}).

\begin{figure*}
\centering
    \includegraphics[width=\columnwidth]{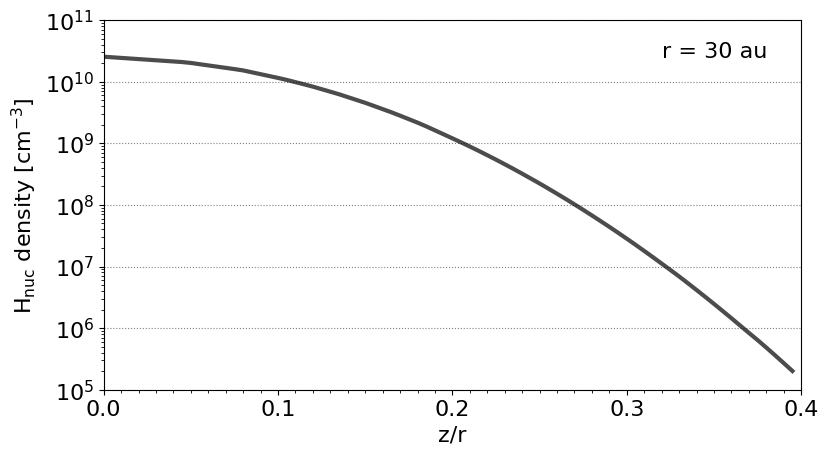}
    \includegraphics[width=\columnwidth]{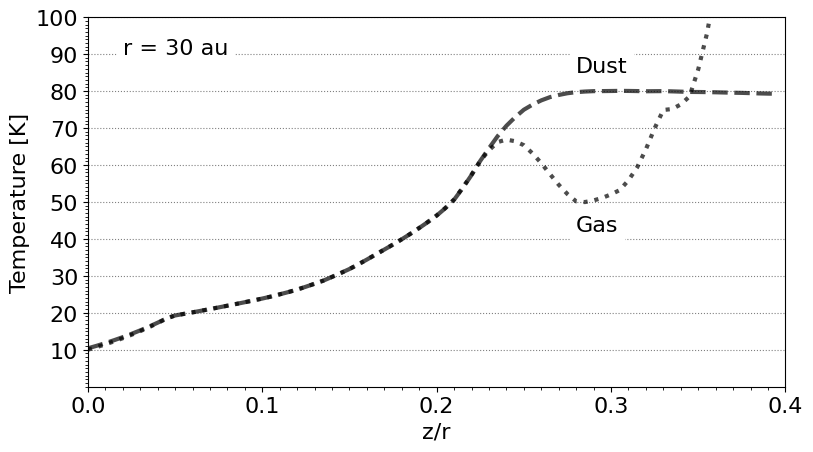}
    \includegraphics[width=\columnwidth]{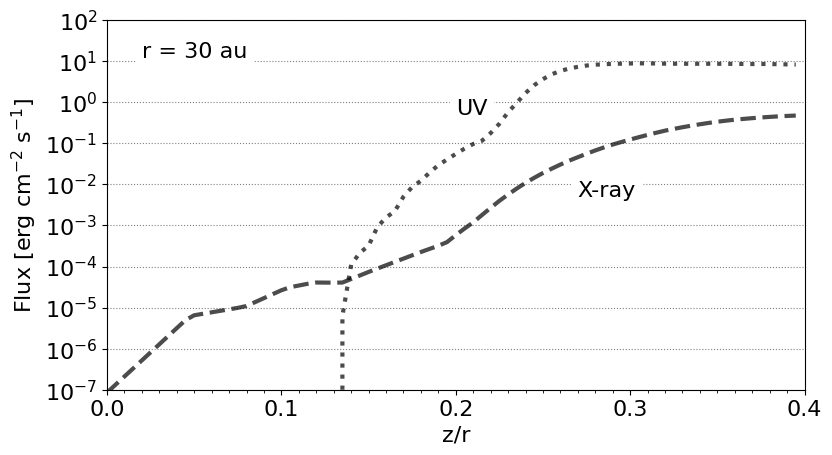}
    \includegraphics[width=\columnwidth]{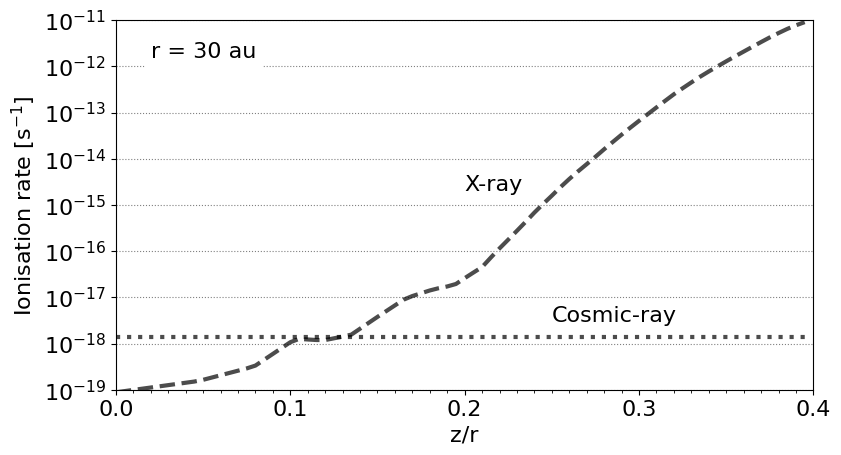}
    \includegraphics[width=\columnwidth]{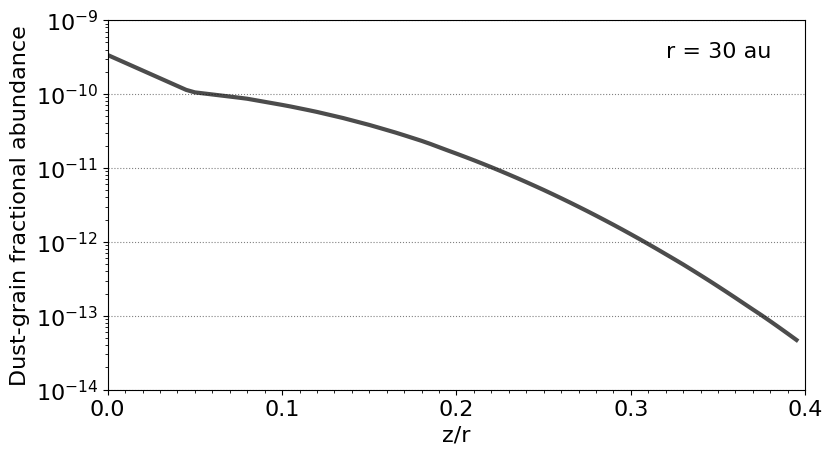}
    \includegraphics[width=\columnwidth]{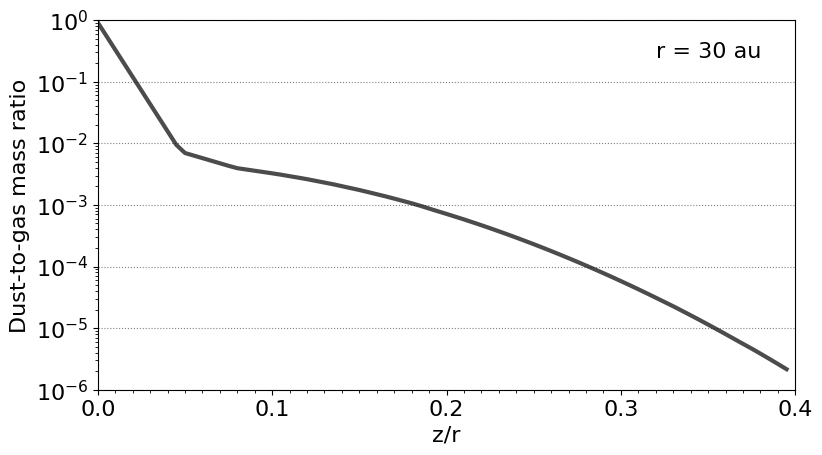}
\caption{Physical conditions across a vertical slice of TW~Hya at a radius of 30~au \citep[model results from][]{calahan21}. Shown are the H$_\mathrm{nuc}$ density (top left), gas (dotted) and dust (dashed) temperature (top right), UV (dotted) and X-ray (dashed) flux (middle left), cosmic-ray (dotted) and X-ray (dashed) ionisation rate (middle right), dust grain fractional abundance (bottom left) and dust-to-gas mass ratio (bottom right), all as a function of $z/r$ (i.e., height divided by radius). \label{twhyaslicephys}}
\end{figure*}

\subsubsection{Physical structure}
Figure~\ref{twhyaslicephys} shows the physical conditions over this slice as a function of height, $z$, over the radius, $r$.  
The hydrogen nuclei density increases from $\sim 10^{5}$~cm$^{-3}$ at the disk ``surface'' ($z/r = 0.4$) to a few times $10^{10}$~cm$^{-3}$ at the disk midplane. 
The gas temperature increases from a value of 10~K at the midplane to  $\gg 100$~K at the disk surface. 
At a height of $z/r \approx 0.23$ the gas and dust temperatures decouple with the gas cooler than the dust up to $z/r \approx 0.35$ beyond which inefficient gas cooling increases the gas temperature relative to that for the dust.  
The dust temperature plateaus at a value of 80~K above $z/r \approx 0.3$ .
The midplane is fully extincted from external UV photons; however, the X-rays are more penetrative and have an appreciable flux of $\sim 10^{-7}$~erg~cm$^{-2}$~s$^{-1}$ at the midplane.  
At a height of $z/r\approx 0.13$ the UV flux becomes larger than that for the X-rays reaching a plateau value of $\sim 10$~erg~cm$^{-2}$~s$^{-1}$ above $z/r \approx 0.26$ which is a factor of $\sim 10^{4}$ times stronger than the average value for the interstellar radiation field.
Despite the X-rays having an appreciable flux in the midplane, below $z/r\approx 0.1$ the ionisation rate is dominated by cosmic rays, and we adopt here the lower-than-canonical ionisation rate  \citep[see, e.g.,][]{Cleeves2015} of $\sim 10^{-18}$~s$^{-1}$.  
Above this height, the X-ray ionisation rate monotonically increases reaching a value of $\sim 10^{-11}$~s$^{-1}$ at $z/r=0.4$. 

The bottom two panels in Figure~\ref{twhyaslicephys} show the dust-grain fractional abundance (left) and dust-to-gas mass ratio (right) as a function of $z/r$ at a radius of 30~au. 
These data include the contributions from both the ``small'' (0.0005 to 1 $\mu$m) and ``large'' (0.005~$\mu$m to 1~mm) dust-grain populations. 
Evident in these plots is the impact of dust settling towards the midplane, with the fractional abundance of dust grains decreasing from a few times 10$^{-10}$ relative to the H nuclei density to a few times 10$^{-14}$ at the top of our disk model.  
These numbers can be compared with the typical fixed value assumed for the interstellar medium of $\sim 10^{-12}$. 
In addition, the dust-to-gas mass ratio is greater than the canonical interstellar value (0.01) below $z/r \approx 0.05$ reaching a maximum of 1 at the disk midplane, and decreasing above this height to reach a minimum of a few times $10^{-6}$ at the disk surface.
This distribution reflects the smaller scale height of the ``large" dust population versus the ``small'' dust population in the best-fit model from \citet{calahan21},  with characteristic scale heights, $h_c$, of 8.4~au and 42~au, respectively, at the characteristic radius, $r_{c}$ of 400~au \citep[see][for full details]{calahan21}. 
The distribution also reflects the different mass contained in the two dust grain populations, with a mass of $1 \times 10^{-4}$~M$_\odot$ in the ``small'' grains and $4 \times 10^{-4}$~M$_\odot$ in the ``large'' grains. 
It is also worth to point out that the best-fit model also suggests a global dust-to-gas mass ratio of 0.02 for TW Hya, which is a factor of 2 larger than that commonly assumed for the ISM (0.01).

\subsubsection{Chemical structure without grain-surface chemistry}
We run 14 iterations of the chemical model, in which we systematically increase the complexity, following a similar approach adopted in \citet[][]{Walsh2014}.  
We do this to separate out and quantify the effects of different processes on the abundance of gas-phase methanol. 
We start with a model which includes gas-phase chemistry, freezeout, and thermal desorption (TD) only, to which we add photodesorption (PD), and/or X-ray desorption (XD), then a model assuming fractal grains (FR) and, finally a model scaling the ice mass reservoir by the dust mass reservoir (D2G).
We run a second series of models with grain-surface chemistry switched on (GR) that includes thermal desorption (TD), photodesorption (PD), and X-ray desorption (XD) by default, and then including reactive desorption (RD), and photoprocessing of the ice (PP), as well as iterations assuming fractal grains (FR), and scaling of the ice mass reservoir (D2G). 
Table~\ref{tab:chemmodels} lists the different model iterations and the chemical processes and assumptions included in each one. 
We take an optimistic approach and extract abundances and column densities at a time of 1~Myr at which we expect most of the chemistry through the disk atmosphere to have reached steady state. 
However, TW Hya has been deemed to be a rather old system, with a recent age estimate of 8~Myr \citep[][]{Sokal2018}.  

\begin{deluxetable*}{ccccccccc}
\tabletypesize{\scriptsize}
\tablecaption{Chemical processes and assumptions included in each iteration of the chemical model.\label{tab:chemmodels}}
\tablehead{
\colhead{Model name} & \colhead{Thermal} & \colhead{Photo-} & \colhead{X-ray }
& \colhead{Grain-surface} & \colhead{Reactive} & \colhead{Ice photo-} & \colhead{Fractal} & \colhead{Scaled} \\ \colhead{} & \colhead{ desorption} & \colhead{desorption} & \colhead{desorption}
& \colhead{chemistry} & \colhead{desorption} & \colhead{
processing} & \colhead{grains} & \colhead{ice mass} }
\startdata
TD              & \checkmark &            & & & & & & \\
TD+PD           & \checkmark & \checkmark & & & & & & \\
TD+XD           & \checkmark &            & \checkmark & & & & & \\
TD+PD+XD        & \checkmark & \checkmark & \checkmark & & & & & \\
TD+PD+XD+FR     & \checkmark & \checkmark & \checkmark & & & & \checkmark & \\
TD+PD+XD+D2G    & \checkmark & \checkmark & \checkmark & & & & & \checkmark \\
TD+PD+XD+FR+D2G & \checkmark & \checkmark & \checkmark & & & & \checkmark & \checkmark \\
GR              & \checkmark & \checkmark & \checkmark & \checkmark & & & & \\
GR+RD           & \checkmark & \checkmark & \checkmark & \checkmark & \checkmark & & & \\
GR+PP           & \checkmark & \checkmark & \checkmark & \checkmark & & \checkmark & & \\
GR+PP+RD        & \checkmark & \checkmark & \checkmark & \checkmark & \checkmark & \checkmark & & \\
GR+RD+FR        & \checkmark & \checkmark & \checkmark & \checkmark & \checkmark & & \checkmark & \\
GR+RD+D2G       & \checkmark & \checkmark & \checkmark & \checkmark & \checkmark & & & \checkmark \\
GR+RD+FR+D2G    & \checkmark & \checkmark & \checkmark & \checkmark & \checkmark & & \checkmark & \checkmark \\
\enddata
\tablecomments{All chemical models include gas-phase chemistry, freeze-out, and thermal desorption by default.}
\end{deluxetable*}

\begin{figure*}
\centering
    \includegraphics[height=5.75cm]{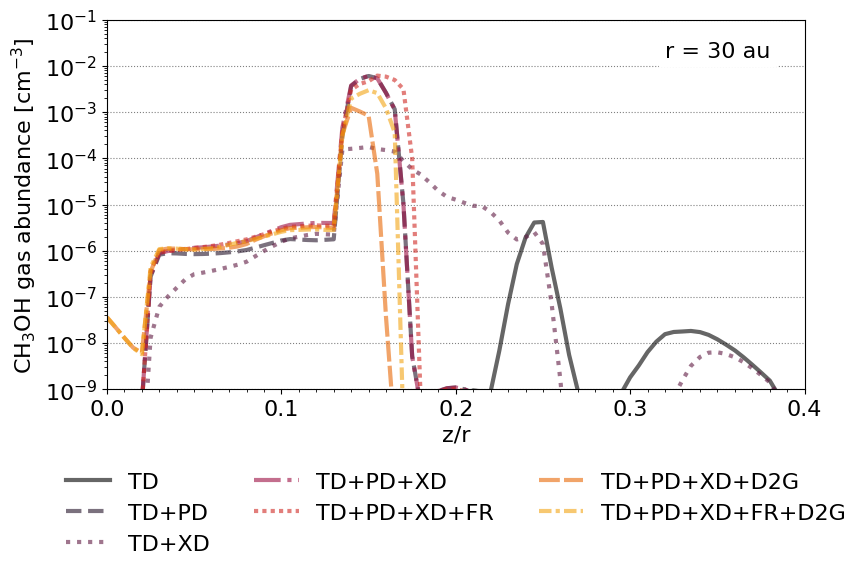}
    \includegraphics[height=5.75cm]{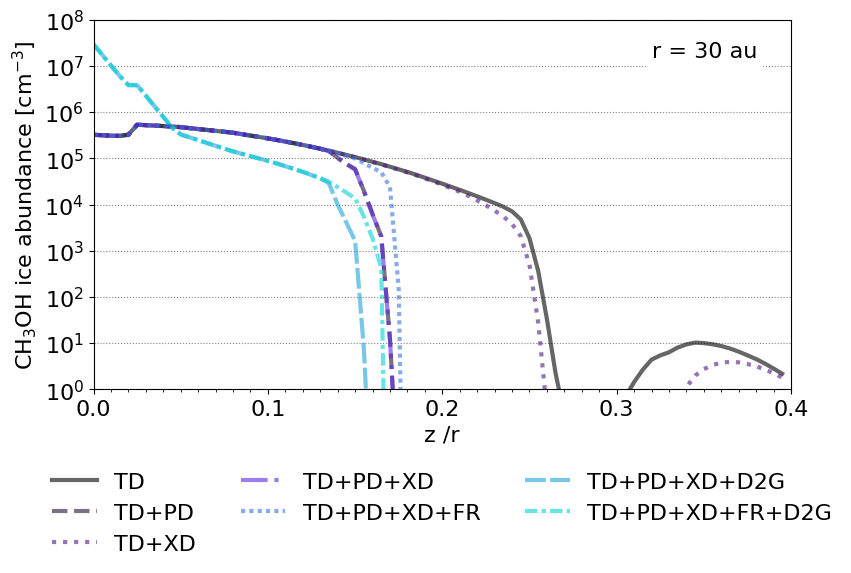}
    \includegraphics[height=5.75cm]{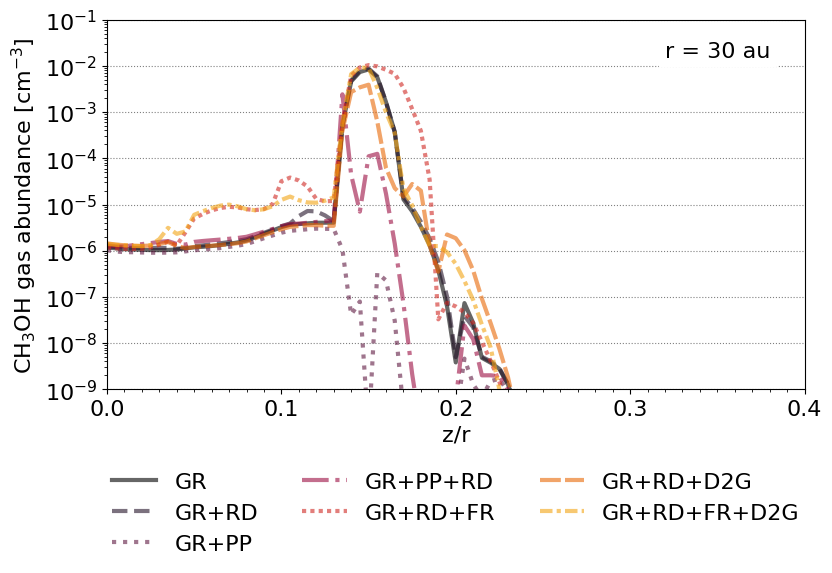}
    \includegraphics[height=5.75cm]{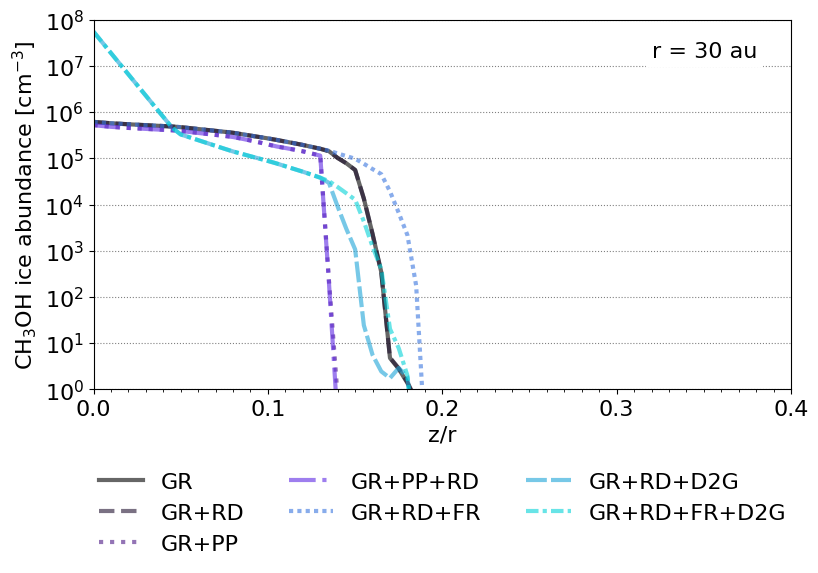}
    \includegraphics[width=\columnwidth]{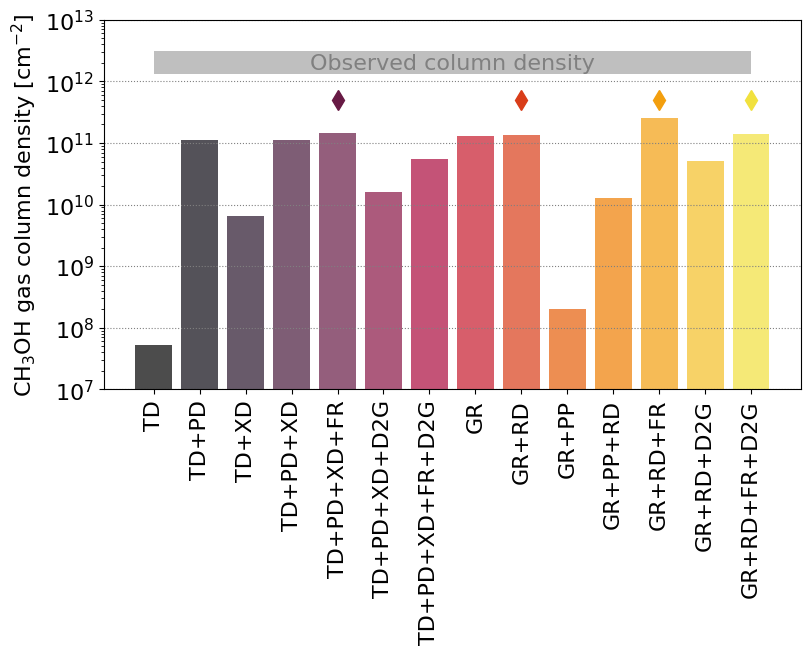}
    \includegraphics[width=\columnwidth]{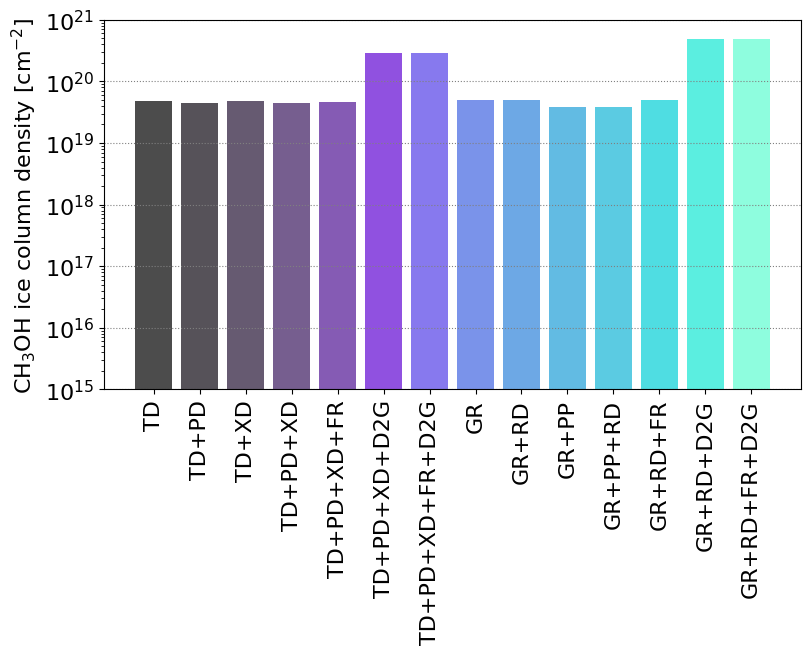}
\caption{Abundance (top and middle; cm$^{-3}$) and column density (bottom; cm$^{-2}$) of methanol gas (left) and ice (right) as a function of $z/r$ (i.e., height divided by radius) at a radius of 30~au. The top row shows results from models with thermal desorption only (TD), with the addition of photodesorption (+PD) and X-ray desorption (+XD), treating the grains as fractal (+FR) and scaling the ice mass by the dust-to-gas mass ratio (+D2G). 
The middle row shows the results from models also including all listed desorption mechanisms and grain-surface chemistry (GR), with the addition of reactive desorption (+RD), photo-processing of the ice (+PP), treating the grains as fractal (+FR), and scaling the ice mass by the dust-to-gas mass ratio (+D2G). 
The disk-averaged column density of gas-phase methanol derived from the observations ($1.8_{-0.5}^{+1.3}\times 10^{12}$~cm$^{-2}$) is shown by the grey box in the bottom left plot. 
The starred models in the bottom left plot are those for which the methanol gas column density is highest and thus for which the full models were run. \label{twhyaslicechem}}
\end{figure*}

Figure~\ref{twhyaslicechem} shows the number density of gas-phase methanol (left) and ice-phase methanol (right) as a function of $z/r$  (top four panels), as well as the total column density of both (bottom two panels) for each iteration of the chemical model.
We choose to show the total number density as opposed to the fractional abundance, as this better reflects where the bulk of the calculated column resides vertically in the disk.
With thermal desorption alone (TD; solid lines in top panels of Fig.~\ref{twhyaslicechem}), the ice reservoir extends to a higher vertical extent ($z/r \lesssim 0.26$) reflecting the cool dust temperature ($\lesssim 80$~K), and the peak in the gas-phase abundance occurs at the methanol snow surface for thermal desorption alone, reaching a value a few times $10^{-6}$~cm$^{-3}$.  
This leads to a column density of $5.2 \times 10^{7}$~cm$^{-2}$ which is around five orders of magnitude lower than the observationally-derived disk-integrated column density of $1.8 \times 10^{12}$~cm$^{-2}$ (see Fig.~\ref{fig:rotational}). 
Further, the gas temperature here is $50 - 70$~K which is higher than the observationally-derived disk-integrated value of 36~K (see Fig.~\ref{fig:rotational}).
Thermal desorption alone cannot reproduced the values derived from the observations. 

The inclusion of photodesorption in the model (TD+PD; dashed lines in top panels of Fig.~\ref{twhyaslicechem}), pushes the position of the methanol ice snow surface down to $z/r \approx 0.17$, where the gas temperature is $30 - 40$~K.
This temperature is more in line with that derived from the observations, and results in a peak abundance of $6 \times 10^{-3}$~cm$^{-3}$ leading to a column density of $1.1 \times 10^{11}$~cm$^{-2}$. 
This is significantly closer to that derived from observations, albeit still an order of magnitude too low.
X-ray desorption alone (XD; dotted lines in top panels of Fig.~\ref{twhyaslicechem}) is less efficient than that for photodesorption and leads to a column density of $6.6 \times 10^{9}$~cm$^{-2}$, although this process does elevate the abundance of gas-phase methanol over a larger vertical extent of the disk. 
With both processes included (TD+PD+XD; dashed-dotted lines in top panels of Fig.~\ref{twhyaslicechem}), photodesorption dominates with very similar profiles and column densities for the results from models TD+PD and TD+PD+XD.

A boost in the dust-grain surface area by assuming fractal grains (TD+PD+XD+FR; tight dotted lines in top panels in Fig.~\ref{twhyaslicechem}), helps to slightly increase the vertical width over which gas-phase methanol reaches its peak abundance relative to the model which does not assume fractal grains. 
This leads to a moderate increase in the column density by $\approx 30$\%.
Scaling of the ice mass and distribution by the dust-to-gas mass ratio (TD+PD+XD+D2G; long dashed lines in top panels of Fig.~\ref{twhyaslicechem}), leads to a drop in the column density by around a factor of $\approx 7$.  
This is because, in this model, the ice is more concentrated in the midplane and depleted in the elevated disk layers where photodesorption is most efficient, leading to a drop in both the peak abundance (to $\sim 10^{-3}$~cm$^{-3}$) and total column density.  
Combining a model with both fractal grains and scaling of the ice mass (TD+PD+XD+FR+D2G; tight dashed-dotted lines in top panels in Fig.~\ref{twhyaslicechem}), leads to an increase in the column density by a factor of $\approx 3$ relative to the model with scaled ice mass alone.
Scaling of the ice-to-gas mass ratio by the dust-to-gas mass ratio has also led to an increase in the column density of methanol ice (bottom-right panel of Fig.~\ref{twhyaslicechem}), from a value of $4.4 - 4.8 \times 10^{19}$~cm$^{-2}$ (with only a small variation between models with different desorption mechanisms included), to $2.9 \times 10^{20}$~cm$^{-2}$.  
This is due to the large increase in abundance of methanol ice in the disk midplane by around two orders of magnitude when this ice mass scaling is applied (see top-right panel of Fig.~\ref{twhyaslicechem}).

So far, the inclusion of photodesorption, and the assumption of fractal grains, has generated the largest calculated column density of gas-phase methanol at 30~au, approaching that derived from the observations. 
We now explore if the inclusion of grain-surface chemistry as well enables a further increase in abundance and column density of gas-phase methanol.

\subsubsection{Chemical structure including grain-surface chemistry}

The middle panels of Fig.~\ref{twhyaslicechem} show the abundances of methanol gas (left) and ice (right) when grain-surface chemistry is switched on.  
As already discussed, methanol is only efficiently formed in the ice phase via hydrogenation of CO ice. 
Hence, this suite of models allow (re)formation of methanol ice following destruction via e.g., photodissociation in the ice and/or gas phase.

The inclusion of grain-surface chemistry alone (GR; solid lines in middle panels of Fig.~\ref{twhyaslicechem}) does not have a major effect on neither the gas-phase nor ice-phase abundance and column density of methanol. 
There is a small increase in the column density of both of $\approx 16$\% and $\approx 12$\%, respectively. 
Adding reactive desorption with a conservative efficiency of 1\% (GR+RD; dashed lines in middle panels of Fig.~\ref{twhyaslicechem}), has a negligible effect, with the results closely overlapping with those including grain-surface chemistry alone.
A model including both grain-surface chemistry and photo-processing of the ices (GR+PP; dotted lines in middle panels of Fig.~\ref{twhyaslicechem}) is catastrophic for the abundance and column density of gas-phase methanol.  
The peak abundance falls to a few times $10^{-6}$~cm$^{-2}$ and the column density falls to a value of $2.0 \times 10^{8}$~cm$^{-2}$.
This likely indicates that adopting the same rates for the photodissociation of ices as those for the gas phase is an oversimplification and/or overestimation in the two-phase model used here, leading to the destruction of methanol ice faster than it can be reformed.
A model including both reactive desorption and photoprocessing (GR+PP+RD; dashed-dotted lines in Fig.~\ref{twhyaslicechem}), does help to boost the abundance of gas-phase methanol relative to the model with photoprocessing only; however, the column density still remains around one order of magnitude lower than that for the base model including grain-surface chemistry (GR).  

The grain-surface model in which reactive desorption and the assumption of fractal grains are also included (GR+RD+FR; tight dotted lines in Fig.~\ref{twhyaslicechem}) leads to the largest column density of gas-phase methanol across all models, $2.5 \times 10^{11}$~cm$^{-2}$, showing that an increase in surface area in conjunction with grain-surface chemistry can help to increase further the abundance of gas-phase methanol.  
This increase is due to the expansion of the vertical width over which methanol reaches its peak abundance of $\sim 10^{-2}$~cm$^{-3}$.  
As found for the models without grain-surface chemistry, scaling of the ice mass by the dust-to-gas mass ratio leads to a boost in the column densities for both gas-phase and ice-phase methanol, relative to the models without this assumption.  
These models predict a further increase in the ice-phase methanol column density to $4.9 \times 10^{20}$~cm$^{-2}$, indicating that grain-surface chemistry is active in producing methanol ice in the disk midplane. 
For gas-phase methanol, the column is increased by a factor of 3.7 when scaling the ice mass, and when fractal grains are not assumed. 
For the converse, scaling of ice mass reduces the column of gas-phase methanol by a factor 1.9.

In summary, photodesorption of methanol ice remains the key process producing methanol in the gas phase.  
Grain-surface chemistry does help to increase the abundance of methanol in the ice phase in the disk midplane, but this has a limited impact on the abundance of gas-phase methanol in the molecular layer. 
The assumption of fractal grains with the addition of grain-surface chemistry and reactive desorption does boost the column density of gas-phase methanol to its highest predicted value at a radius of 30~au; however, this is still around an order of magnitude lower than the disk-integrated value derived from observations. 

\subsection{Abundance of methanol gas and ice in the full model}

We next explore the two-dimensional abundance of gas-phase and ice-phase methanol as well as the radially-dependent column densities. 
We do this for the sub-set of models which predict the highest column density of gas-phase methanol at 30~au. 
These are identified with a diamond symbol in the bottom-left panel of Fig.~\ref{twhyaslicechem}. 
These are models with i) thermal desorption, photodesorption, X-ray desorption and fractal grains (TD+PD+XD+FR), ii) the same as model i) but including grain-surface chemistry and reactive desorption (GR+RD), iii) the same model as ii) but including fractal grains (GR+RD+FR), and iv) the same model as iii) but scaling the ice mass by the dust-to-gas mass ratio (GR+RD+FR+D2G).

\subsubsection{Physical structure of the full model}

\begin{figure*}
\centering
    \includegraphics[width=\columnwidth]{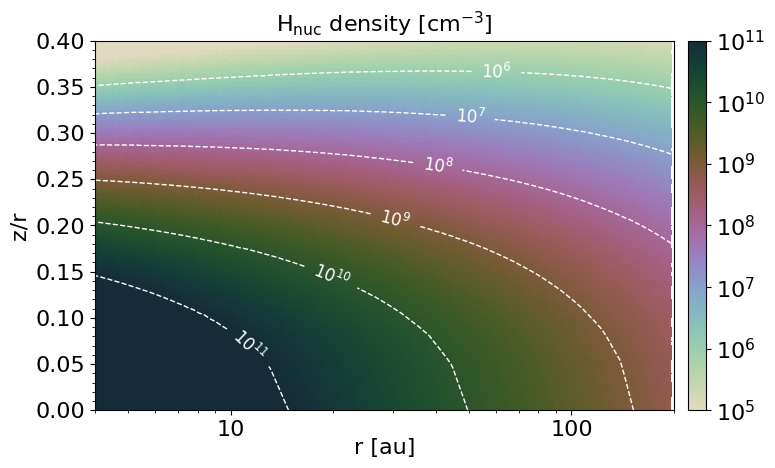}
    \includegraphics[width=\columnwidth]{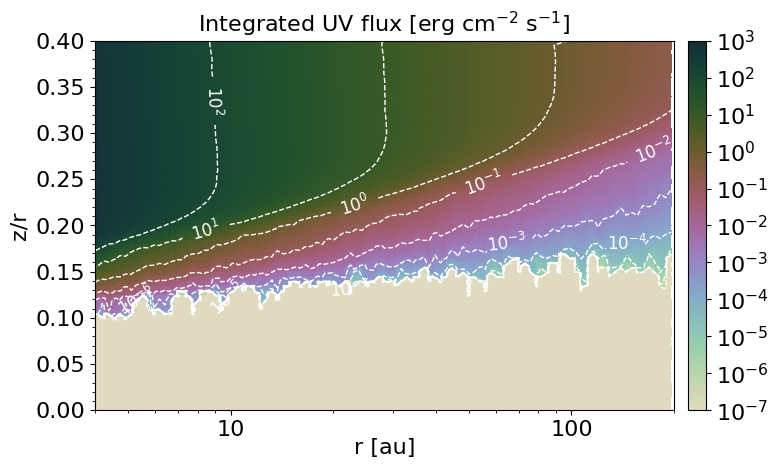}
    \includegraphics[width=\columnwidth]{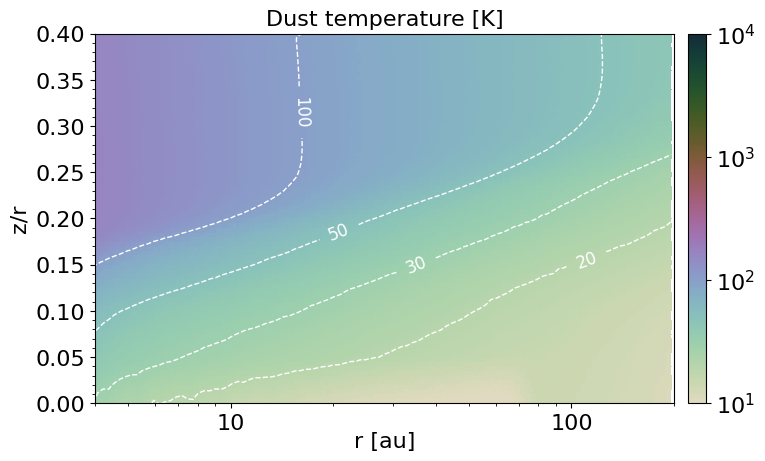}
    \includegraphics[width=\columnwidth]{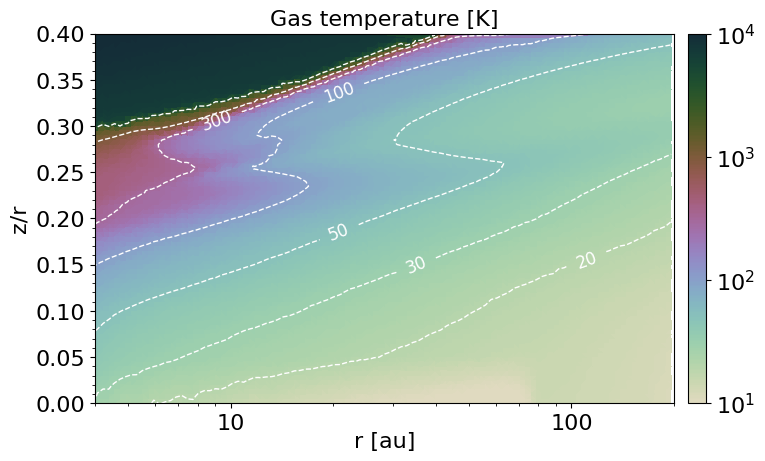}
    \includegraphics[width=\columnwidth]{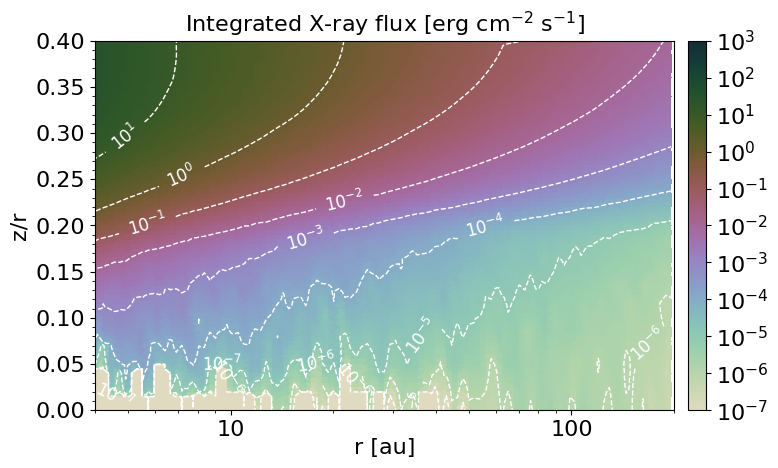}
    \includegraphics[width=\columnwidth]{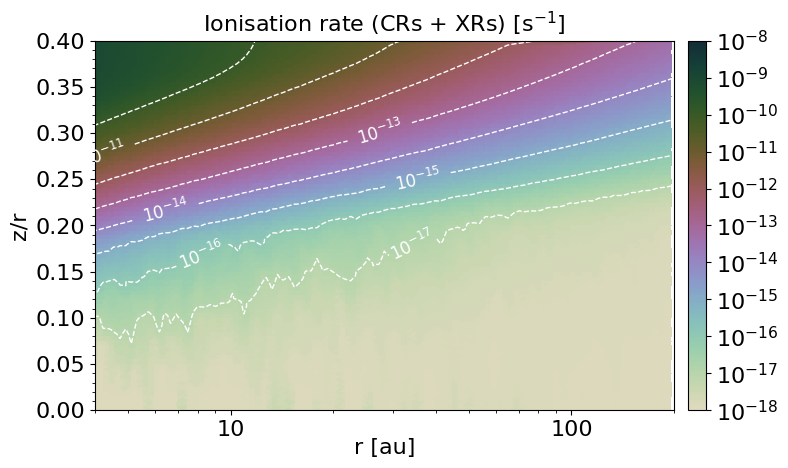}
    \includegraphics[width=\columnwidth]{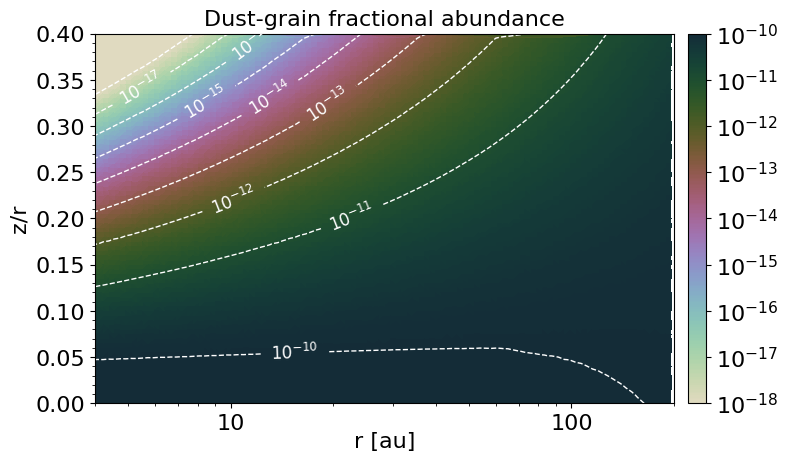}
    \includegraphics[width=\columnwidth]{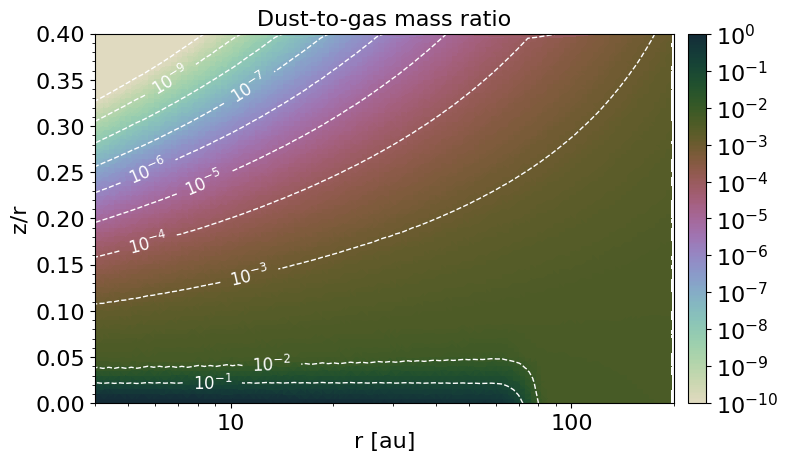}
\caption{Physical conditions as a function of disk radius, $r$ and height divided by the radius, $z/r$ (model results from \citealt{calahan21}). 
Shown are the H$_\mathrm{nuc}$ density (top left), integrated UV flux (top right), dust temperature (second left), gas temperature (second right), integrated X-ray flux (third left), total ionisation rate (third right), dust grain fractional abundance (bottom left), and dust-to-gas mass ratio. \label{twhyafullphys}}
\end{figure*}

Figure~\ref{twhyafullphys} shows the full two-dimensional structure of the disk for TW~Hya from \citet{calahan21}. 
Shown are the H nuclei density (top left), integrated UV flux (top right), dust temperature (second left), gas temperature (second right), integrated X-ray flux (third left), total ionisation rate (third right), dust grain fractional abundance (bottom left), and dust-to-gas mass ratio (bottom right).  
The maps for the UV and X-ray integrated fluxes contain noise related to the stochastic method used to compute the UV and X-ray radiation fields in rac-2d \citep[see][and references therein]{Du2017}.
The structure follows a similar behaviour to that already described for the slice of the disk at 30~au.  
The disk midplane is well shielded from UV radiation below a height of $z/r$ from 0.1 (at 4au) to 0.15 (at 200 au).  
The gas and dust temperatures decouple above a height of $z/r$ of 0.2 (at 4 au) to 0.3 (at 200 au).
X-rays are able to penetrate deeper into the disk than UV photons leading to an appreciable X-ray flux in the disk midplane ($\gtrsim 10^{-7} - 10^{-6}$~erg~cm$^{-2}$~s$^{-1}$); however, cosmic rays still dominate the ionisation rate below a height of $z/r$ from 0.05 (at 4 au) to 0.2 (at 200 au). 
The impact of dust settling and drift is clearly seen in the bottom two panels of Fig.~\ref{twhyafullphys}. 
The fractional abundance of grains is $\gtrsim 10^{-10}$ relative to the H nuclei density below $z/r$ equal to 0.05 out to a radius of $\approx 60$~au, leading to a higher dust-to-gas mass ratio in this region than commonly assumed in the ISM (i.e., $>0.01$), and a lower dust-to-gas mass ratio everywhere else in the disk.

\subsubsection{Chemical structure in the full models}

\begin{figure*}
\centering
    \includegraphics[width=\columnwidth]{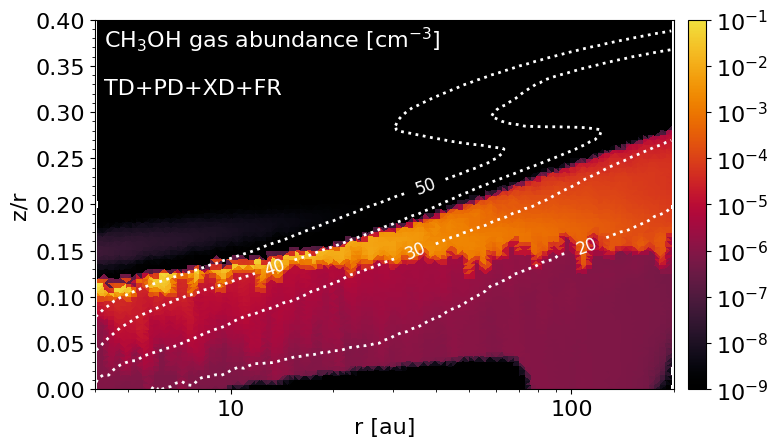}
    \includegraphics[width=\columnwidth]{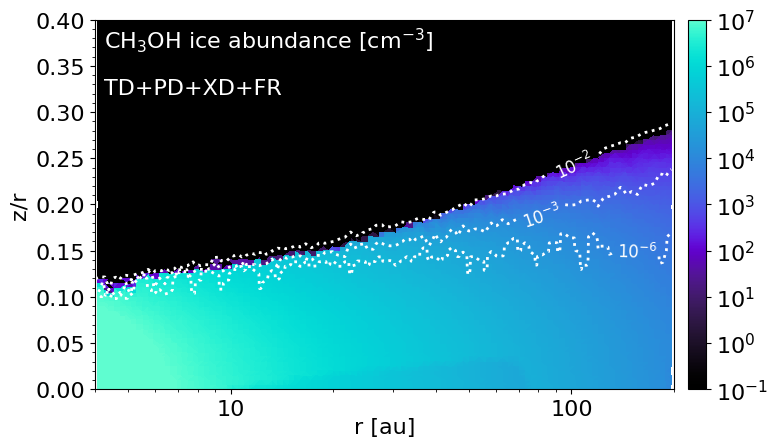}
    \includegraphics[width=\columnwidth]{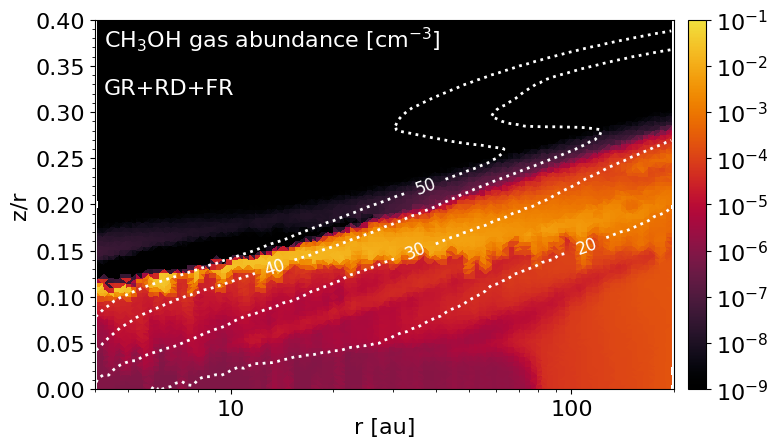}
    \includegraphics[width=\columnwidth]{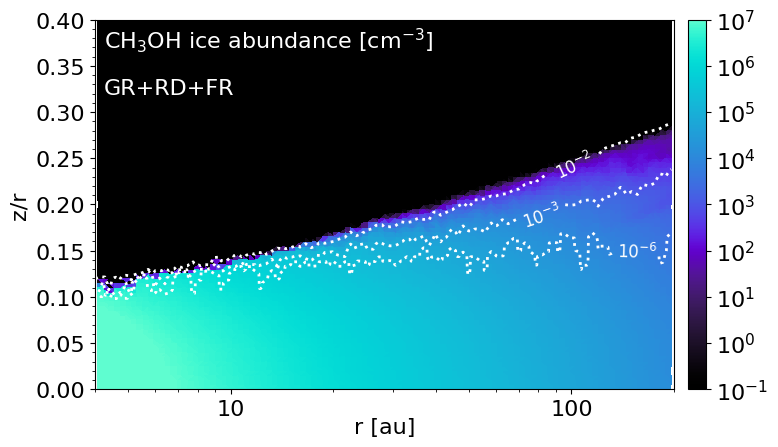}
    \includegraphics[width=\columnwidth]{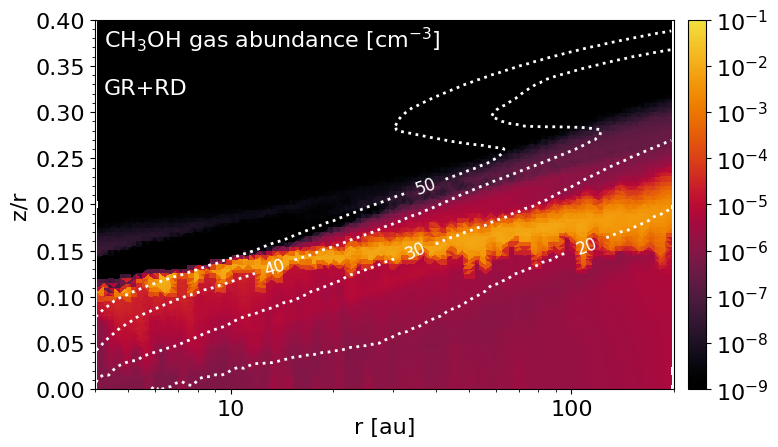}
    \includegraphics[width=\columnwidth]{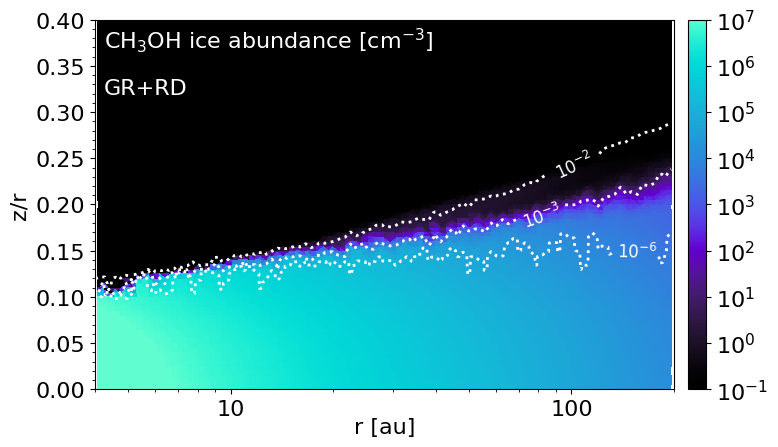}
    \includegraphics[width=\columnwidth]{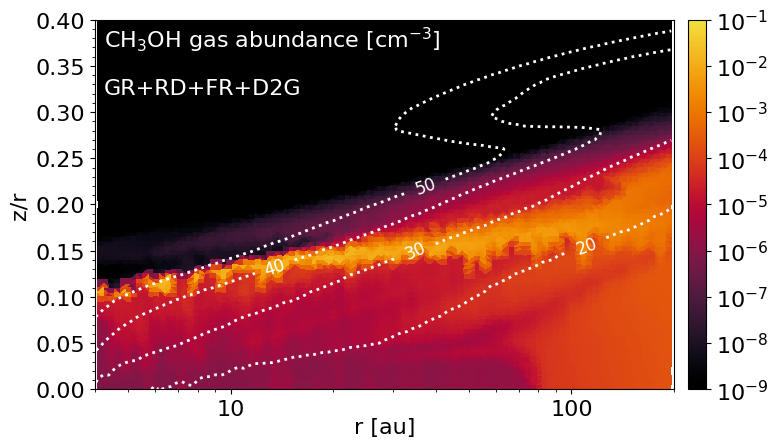}
    \includegraphics[width=\columnwidth]{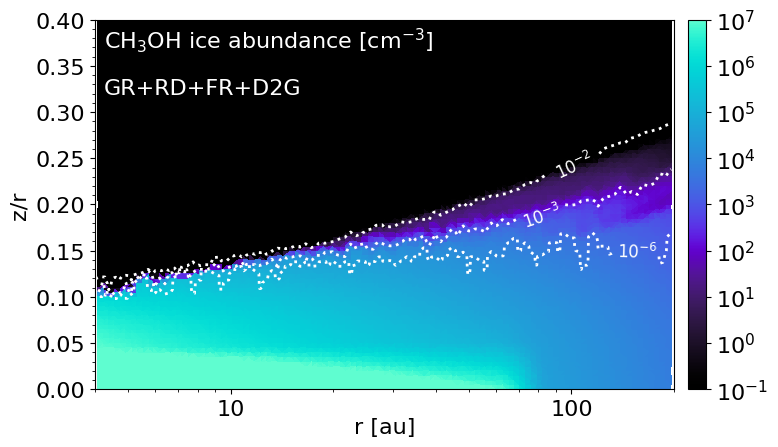}
    \caption{Number density of gas-phase (left) and ice-phase (right) methanol as a function of disk radius, $r$, and height divided by the radius, $z/r$. Shown are the results from the four models which predict the highest column density of gas-phase methanol at 30~au: TD+PD+XD+FR (top), GR+RD+FR (second), GR+RD (third) and GR+RD+FR+D2G (bottom). 
    The contours in the left-hand plots present the gas temperature structure in units of K, whereas those in the right-hand plots present the integrated UV field in units of erg~cm$^{-2}$~s$^{-1}$.  
    All model ingredients are listed in Table~\ref{tab:chemmodels}. \label{twhyfullchem}}
    \end{figure*}

Figure~\ref{twhyfullchem} shows the two-dimensional number densities (cm$^{-3}$) of gas-phase (left) and ice-phase (right) methanol for the four models which predict the highest column density of gas-phase methanol at 30~au. 
The gas-phase methanol is most abundant in the warm molecular layer spanning a gas temperature of $\approx 40$ to 50 K in the inner disk ($\lesssim 10$~au) to a gas temperature range of $\approx 20$ to $\approx 30$ to 40~K (depending on the model) in the outer disk ($\approx 100$~au).  
These temperature ranges overlap with that derived from the observations of gas-phase methanol in TW Hya. 
In the model which includes thermal and non-thermal desorption and fractal grains but excludes grain-surface chemistry (TD+PD+XD+FR; top panels in Fig.~\ref{twhyfullchem}), the location of the gas-phase methanol is bounded below and above by integrated UV field strengths of $\sim 10^{-6}$~erg~cm$^{-2}$~s$^{-1}$ and $\sim 10^{-2}$~erg$^{-2}$~s$^{-1}$, respectively. 
This indicates the sweet spot over which photodesorption is efficient at releasing ices from the dust grains, but at which photodissociation is not yet efficient at destroying these species in the gas phase.
All models have this same lower boundary with some small variations in upper boundary depending on the model. 
In the bottom two models shown in Fig.~\ref{twhyfullchem} (GR+RD and GR+RD+FR+D2G), this upper boundary lies at the lower integrated UV flux of $\sim 10^{-3}$~erg$^{-2}$~s$^{-1}$, indicating the importance of the available dust grain surface area and initial ice abundance in setting the position of the snow surface. 

In the models with both grain-surface chemistry and assuming fractal grains (GR+RD+FR and GR+RD+FR+D2G), there is a region in the outer disk ($\gtrsim 60$~au) extending towards the midplane over which gas-phase methanol also reaches an appreciable abundance ($\sim 10^{-3}$~cm$^{-3}$). 
The assumption of fractal grains, with the corresponding increase in available dust-grain surface area, is helping to increase the efficiency of grain-surface chemistry and non-thermal desorption in this region.
In the model in which the ice mass has been scaled by the dust-to-gas mass ratio (GR+RD+FR+D2G; bottom panels of Fig.~\ref{twhyfullchem}), the effects of the redistribution of ices can be seen in the methanol ice abundance.  
There is an increase in the abundance within the disk midplane ($r \lesssim 70$~au; $z/r \lesssim 0.05$) corresponding to where the dust-to-gas mass ratio is greater than 0.01, relative to the models without this scaling. 

\begin{figure*}
\centering
    \includegraphics[width=\columnwidth]{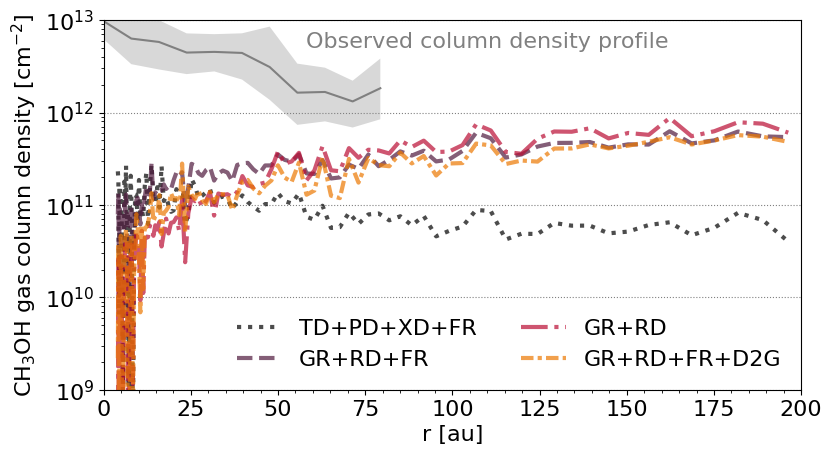}
    \includegraphics[width=\columnwidth]{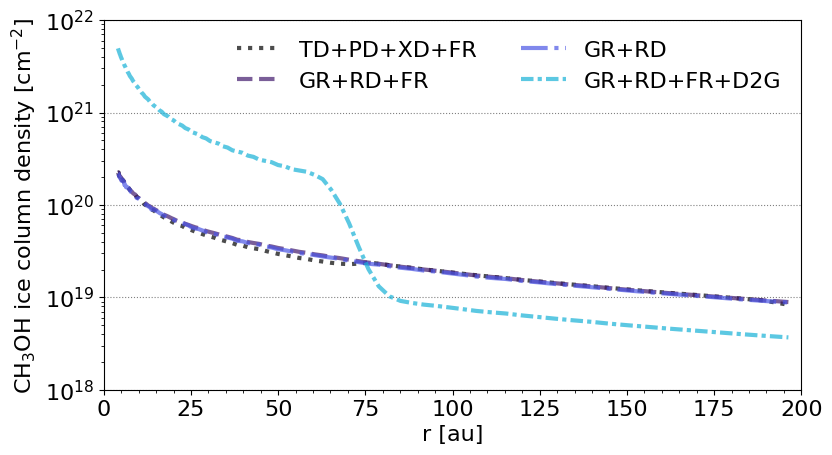}
    \includegraphics[width=\columnwidth]{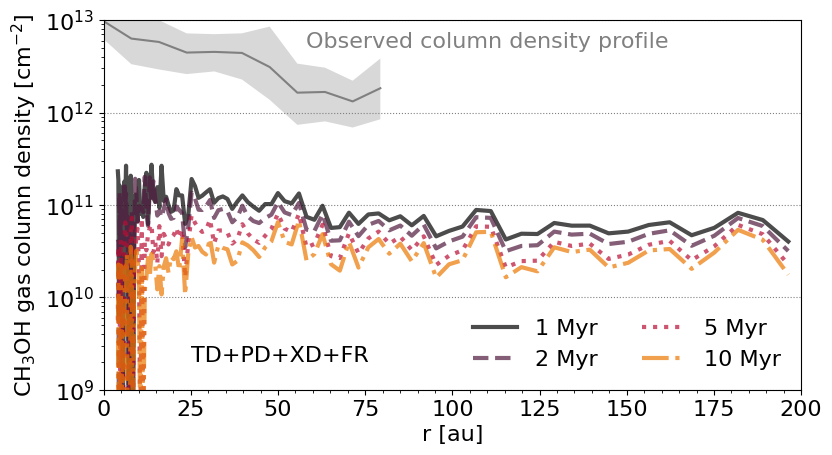}
    \includegraphics[width=\columnwidth]{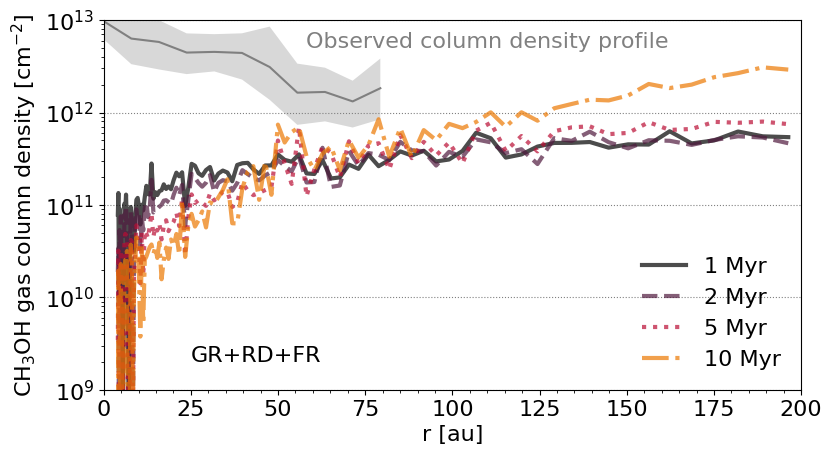}
    \caption{Column density of gas-phase (top left) and ice-phase (top right) methanol as a function of disk radius, $r$, at a time of 1~Myr. 
    Shown in the top panels are the results from the four models which predict the highest column density of gas-phase methanol at 30 au: TD+PD+XD+FR (dotted lines), GR+RD+FR (dashed lines), GR+RD (dotted-dashed lines) and GR+RD+FR+D2G (tight dotted-dashed lines). All model ingredients are listed in Table~\ref{tab:chemmodels}. 
    Shown in the bottom panels are the results from two models, one without (TD+PD+XD+FR; bottom left) and with (GR+RD+FR; bottom right) grain-surface chemistry and reactive desorption, at times of 1, 2, 5, and 10~Myr.
    The grey shaded region in the gas-phase column density plots shows the column density profile inferred from the observations (see Fig.~\ref{fig:rotational}). 
    \label{twhyfullcol}}
    \end{figure*}

The top panels of Fig.~\ref{twhyfullcol} show the vertically-integrated column density of gas-phase (left) and ice-phase (right) methanol as a function of radius, $r$, for the four models discussed here.  
The three models which include grain-surface chemistry (GR+) all have a very similar column density profile for gas-phase methanol which increases from 1 -- 2 $\times~10^{11}$~cm$^{-2}$ at 30~au to 5 -- 6 $\times~10^{11}$~cm$^{-2}$ at 200~au.  
In contrast, the model without grain-surface chemistry (TD+PD+XD+FR; dotted lines in Fig.~\ref{twhyfullcol}) has a column density profile which generally decreases with radius, from a value of $1.5 \times 10^{11}$~cm$^{-2}$ at 30~au to $4.0 \times 10^{10}$~cm$^{-2}$ at 200 au.  
It is worth to note that no model reaches the disk-integrated column density derived from the observations ($1.8 \times 10^{12}$~cm$^{-2}$) by a time of 1~Myr with the models including grain-surface chemistry achieving the highest column densities, albeit in the very outer disk and still a factor of 2 -- 3 lower than the column density derived at $\approx 75$~au. 
Arguably, the model without grain-surface chemistry has a column density profile more similar to the shape the profile constrained from the observations (see Fig.~\ref{fig:rotational_radial}); however, this would need to be scaled up by a factor of $\sim 100$ within 50~au and a factor of $\sim 10$ beyond in order to be fully consistent.  

Regarding the ice column density, all models with the exception of that including the scaling of the ice mass, have the same or very similar profiles. 
For the model in which the ice mass is scaled by the dust-to-gas mass ratio (GR+RD+FR+D2G; tight dotted-dashed lines in Fig.~\ref{twhyfullcol}), the ice column density is greater by a factor of $\approx 25$ within 60~au, and is lower by a factor of $\approx 2.5$ beyond 80~au.\
This reflects the distribution of dust grains shown in the bottom right panel of Fig.~\ref{twhyafullphys}.

All model results discussed thus far are based on abundances extracted at a time of 1~Myr.  This is based on previous gas-grain models that suggested that methanol ice, and as a consequence methanol gas, can be depleted over longer timescales \citep[$\sim$ 8~Myr;][]{Ligterink2018} which would be more akin to the age of TW Hya.  
In the bottom panels of Fig.~\ref{twhyfullcol} we show the column density of gas-phase methanol as a function of radius, $r$, for two models, one without (left) and with (right) grain-surface chemistry and reactive desorption, at timesteps of 1, 2, 5, and 10 Myr.  
We choose these models to assess the impact of grain-surface chemistry and reactive desorption on the column density of gas-phase methanol predicted through the disk.
For the model without grain-surface chemistry and reactive desorption, the column densities are reduced only by a factor of a few over timescales from 1 to 10~Myr.
For the model including these processes, there is little difference in the results from 1 to 5~Myr, however, by 10 Myr, there is an appreciable increase in the column density of gas-phase methanol in the outer disk to $\sim 10^{13}$~cm$^{-2}$.  
This shows that grain-surface chemistry can help to boost the abundance of gas-phase methanol in the outer disk, but only on long timescales.

\section{Discussion}
\label{sec:discussion2}

What have we learned from the chemical models? The model results confirm photodesorption as a key mechanism releasing methanol from the ice into the gas phase.  
This is in agreement with previous gas-grain chemical modelling of the disk of TW Hya \citep{Parfenov2017,Walsh2018}. 
The assumption of fractal grains, and the inclusion of grain-surface chemistry and reactive desorption, can help to boost further the column density of gas-phase methanol.  
However, reactive desorption tends to increase the abundance of gas-phase methanol deeper in the disk where the gas temperature is $\lesssim 20$~K, rather than in the molecular layer (see also \citealt{Parfenov2017} and \citealt{Walsh2018}). 
On the other hand, enabling photodissociation of the ice as well as scaling the ice mass by the dust-to-gas mass ratio, generally have the effect of decreasing the column density of methanol in the molecular layer of the disk, relative to the models excluding these ingredients.  
In the former case, methanol ice is destroyed more rapidly than it can be reformed, and in the latter case, the scaling reduces the initial abundance of methanol ice in the layers of the disk exposed to sufficient UV radiation to drive photodesorption. 

The full chemical models predict that gas-phase methanol resides in a layer with a temperature range consistent with that derived from the observations; however, most models are not able to reach the disk-integrated column density of methanol derived from the observations. 
The one exception to this is the model with grain-surface chemistry, reactive desorption and fractal grains, which reaches a column density of $3 \times 10^{12}$~cm$^{-2}$ similar to that inferred from the observations, albeit only at the very outermost radius of the disk at a time of 10~Myr (see Fig.~\ref{twhyfullcol}). 
This timescale is more in line with the age of the TW Hya system; hence, more efficient grain-surface chemistry in the inner disk regions (e.g., due to a larger dust-grain surface area and/or more efficient diffusion and reactive desorption) may help to boost the gas-phase methanol abundance in the inner disk.

The {\em radial distribution} of gas-phase methanol predicted by the models including grain-surface chemistry (as traced by the column density) differs from that inferred from the observations. 
The models suggest that the column density of gas-phase methanol monotonically increases with radius, whereas the observations suggest the opposite.  
This is again similar to earlier gas-grain models for TW~Hya \citep[][]{Parfenov2017,Walsh2018}. 
Even in models where we have reduced significantly the initial abundance of ice phase methanol due to scaling of the initial abundance by the dust-to-gas mass ratio, there appears to remains sufficient dust grain surface area to form methanol ice in the disk on timescales of $\sim 10$~Myr (see Fig.~\ref{twhyfullcol}). 

So how do we reconcile the difference between model predictions and observations in terms of gas-phase methanol column density and distribution?  
One limitation of our chemical models is that they are static in nature and do not include the effects of vertical mixing driven by turbulent diffusion.  
Generic models of a protoplanetary disk (i.e., not tailored to TW Hya) which include turbulent mixing predict that the abundance (and thus column density) of gas-phase methanol can increase by factors of $\sim 100$ in the molecular layer of disk, for the case of strong/fast mixing \citep{Semenov2011,Heinzeller2011,Furuya2014,Parfenov2016}.  
Hence, the inclusion of turbulent mixing may increase the column density to values in better agreement with observations.  
However, measurements of the turbulence in TW Hya have suggested that it is sub-sonic, with estimates for the turbulent velocity, $\lesssim 0.1~c_s$, where $c_s$ is the local sound speed in the disk \citep{Flaherty2018,Teague2018}.
This turbulent velocity corresponds to $\alpha \lesssim 0.01$ where $\alpha$ is the commonly assumed scaling parameter used to estimate the turbulent viscosity in accretion disks \citep[$\nu \approx \alpha c_s H$, where $H$ is the scale height of the disk;][]{Shakura1973}. 
Hence, it is likely that vertical mixing is weak in TW Hya.  However, determining the precise impact of vertical mixing on the abundance and distribution of gas-phase methanol will require a disk-specific chemical model of TW Hya that self-consistently includes turbulent mixing.

\begin{figure*}
    \centering
    \includegraphics[width=\columnwidth]{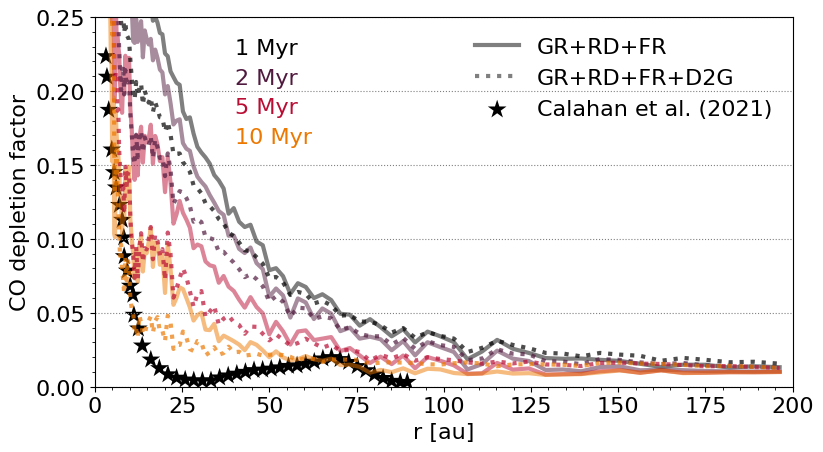}
    \includegraphics[width=\columnwidth]{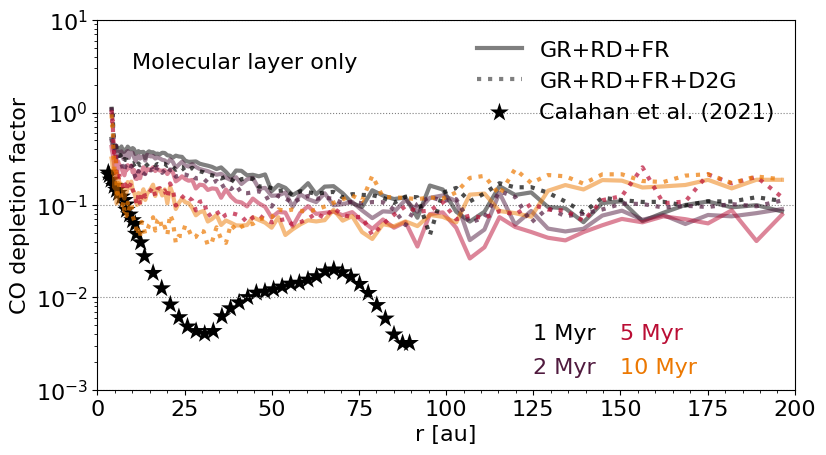}
    \caption{CO depletion factor as a function of radius, $r$, at 1, 2, 5, and 10~Myr for the model that includes grain-surface chemistry, reactive desorption, and fractal grains (GR+RD+FR; solid lines), and the same model but also scaling the ice mass by the dust-to-gas mass ratio (GR+RD+FR+D2G; dotted lines).
    The empirically-derived CO depletion factors from \citet{calahan21} are overlaid in black stars. 
    The left-hand panel shows the ratio of gas-phase CO to \ce{H2} for the full vertical extent of the disk, whereas the right-hand plot shows that calculated for the molecular layer only (i.e., restricting the calculation to above the CO snow surface only).}
    \label{fig:CO-to-H2}
\end{figure*}

In addition, in our models with dust-to-gas mass ice mass scaling, we only rescale the initial ice abundances.  
There is now compelling observational evidence that gas-phase CO is depleted by a factor of up to $\sim 100$ in TW Hya \citep[see, e.g.,][]{Zhang2017, calahan21}. 
This is thought to arise via a combination of chemistry and dust evolution, with chemistry helping to convert CO to less volatile forms which freeze out as ice, and dust evolution helping to leach this converted CO from the outer and upper regions of the disk, concentrating it in the inner midplane \citep[e.g.,][]{booth_2017,Schwarz2018,Bosman2018,booth_2019,Krijt2020}. 
Our simulations begin with an assumed fractional abundance for CO (gas plus ice) of $7.4 \times 10^{-5}$ with respect to the H nuclei density, so we begin with a very moderate depletion only of CO (by a factor of $\approx 2$ compared with assuming that all carbon is contained in CO). 
An excess of CO available in the outer disk could be helping to enhance the formation and survival of methanol ice and gas in the outer disk, thus boosting the column density there, and creating the column density profile in conflict with that derived from observations.  
In Fig.~\ref{fig:CO-to-H2} we present a calculation of the CO depletion factor as a function of radius, $r$, at 1, 2, 5, and 10 Myr.  
We show this for the models with grain-surface chemistry, reactive desorption, and fractal grains, and with and without scaling of the ice mass by the dust-to-gas mass ratio. 
We choose these models to assess the impact of scaling of the ice mass on the CO depletion factor.
The CO depletion factor here is calculated by taking the ratio of the vertically-integrated CO gas-phase column density to the \ce{H2} column density, and then scaling this with the total abundance of carbon in the disk with respect to \ce{H2} ($2.8\times 10^{-4}$).  
We show two versions of this plot, one for the full vertical range of the disk (left-hand panel), and one where we restrict the calculation to above the CO snow surface (right-hand panel). 

The left-hand panel in Fig.~\ref{fig:CO-to-H2} shows that the depletion of CO over the full vertical column increases with time \citep[in agreement with previous gas-grain models, e.g.,][]{Bosman2018}, and that the model with ice mass scaling also generally has higher CO depletion factors than that without.  
This CO depletion factor falls below 0.1 (i.e., 10 times less CO than canonical) beyond $\approx 70$, $\approx 60$, $\approx 50$, and $\approx 30$~au at times of 1, 2, 5, and 10~Myr, respectively in the model without ice-mass scaling.  
These values are similar at 1 and 2 Myr for the model with ice mass scaling; however, at 5 and 10~Myr, this model falls below a value of 0.1 at a radius of $\approx 35$ and $\approx 10$~au, respectively.
All models tend towards a similar value of $\sim 0.01$ in the very outer disk, which is consistent with the factor of $\sim 100$ depletion inferred from observations.  
Note that this depletion level has naturally emerged from a model that includes only static chemical evolution with no dynamic element to the dust population.  
If we restrict the calculation to above the CO snow surface and in the molecular layer only, the maximum CO depletion factor reached by both models at all times tends to $\sim 0.1$ with a non-linear behaviour with time indicating a recovery in the CO abundance in the molecular layer at timescales of 10~Myr (yellow lines in Fig.~\ref{fig:CO-to-H2}).  

We can compare the profiles in Fig.~\ref{fig:CO-to-H2} with radially-dependent depletion factors empirically determined for TW Hya from observations in \citet[][plotted as black stars in Fig.~\ref{fig:CO-to-H2}]{calahan21}.  
In the inner disk, we get best agreement for the model with ice mass scaling at time of 10~Myr which has the steepest depletion profile with radius. 
However, we do not recover the very low depletion factors inferred from observation for the outer disk ($\ll 0.01$), indicating that additional processes, such as dust evolution, are still needed.  
Inspection of these model results suggest that imposing further CO depletion in the outer disk and/or including dust evolution in a 2D model with gas-grain chemistry, could help to reduce the column density of methanol in the outer disk and bring the shape of the modelled column density profile more in line with that inferred from the observations.

\begin{figure}
    \centering
    \includegraphics[width=\columnwidth]{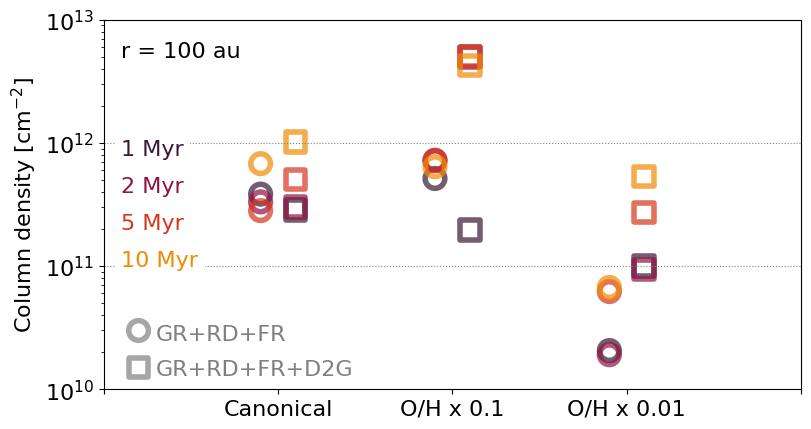}
    \caption{Column density of gas-phase methanol (\ce{CH3OH}) at a radius of 100~au for the canonical elemental abundances (C/O = 0.43; O/H = $3.2\times 10^{-4}$), and two models where oxygen is globally depleted by factors of 10 and 100 and C/O is set to 2.  Shown are the results for the models with grain-surface chemistry, reactive desorption and fractal grains and with (GR+RD+FR+D2G) and without (GR+RD+FR) ice-to-gas mass scaling.}
    \label{fig:CH3OH_depleted}
\end{figure}

We run a final set of models where we apply global oxygen depletion by factors of 10 and 100 and impose a C/O ratio of 2 in both cases \citep[as suggested in previous work;][]{Bergin2016,Zhang2019,calahan21,Cleeves2021}.  
We do this for a vertical slice of the disk at 100 au for the two models for which we have computed the CO depletion factor generated by chemistry alone (GR+RD+FR and GR+RD+FR+D2G).  
Fig.~\ref{fig:CH3OH_depleted} shows the vertically-integrated column density for gas-phase methanol as a function of time (indicated by the colour scale), for each set of elemental abundances and each model.  
Somewhat counter intuitively, global depletion of oxygen by a factor of 10 and a carbon-rich environment does not lead to a drop in the column density of gas-phase methanol. 
Under these conditions, the models with ice-to-gas mass scaling generate an increase in the column density by almost an order of magnitude at timescales of 2~Myr and longer, relative to the canonical model with no imposed depletion. 
In the carbon-rich environment almost all oxygen will be in the form of CO; hence, methanol ice (and by extension gas) can still form via hydrogenation of CO ice, with little to no competition from hydrogenation of oxygen to form water ice. 
The model with ice-mass scaling applied has a higher C/O ratio relative to that without because dominant oxygen carriers in the ice such as \ce{H2O} and \ce{CO2} are further depleted by this scaling relative to CO (recall that we begin with 50\% only of CO in ice form). This is why the column densities for model GR+RD+FR+D2G are generally larger than those from model GR+RD+FR.

We do see a drop in the column density of methanol to $\lesssim 10^{11}$~cm$^{-2}$ when oxygen is depleted globally by a factor of 100, but only for the model where we do not scale the ice mass by the dust mass (GR+RD+FR; see explanation above). 
These values are more consistent with the observations of gas-phase methanol analysed here and suggest a radial gradient of C/O across the disk.
Hence, the lack of methanol emission from the outer disk supports the requirement for oxygen (and by extension CO) to be depleted by around two orders of magnitude in the outer disk of TW Hya, as suggested in the analysis of emission from other carbon-bearing species in the outer disk of TW Hya with ALMA \citep[e.g.,][]{Bergin2016,Cleeves2021}. 

\section{Summary}
\label{sec:summary}

We report the successful detection of individual rotational lines of gas-phase methanol in the disk around TW Hya with ALMA. 
We detect emission from four lines spanning upper level energies from 16 to 38~K.  
A rotational diagram analysis constrains the disk-integrated column density to $1.8^{+1.3}_{-0.5}\times 10^{12}$~cm$^{-2}$ and disk-averaged rotational temperature to $35.9^{+25.9}_{-10.6}$~K.  
A radially resolved analysis confirms that the emission is compact, peaking within the spatial extent of mm-sized grains.  
The radially-resolved rotational temperature of gas-phase methanol is similar to that derived from the disk-averaged analysis and appears constant over the disk.  
The radially-resolved column density peaks at $10^{13}$~cm$^{-2}$ in the inner disk, and monotonically decreases to a few times $10^{12}$~cm$^{-2}$, i.e., to a similar value as found in the disk-integrated analysis.

A comparison with the outputs of gas-grain chemical models of the disk of TW Hya confirms photodesorption as a key process releasing methanol from the ice mantle. 
This is despite adopting rates which include fragmentation upon photodesorption as confirmed in laboratory experiments.
Models including photodesorption predict that the gas-phase methanol resides in a layer in the disk at a temperature of 20 to 50 K, consistent with the rotational temperature constrained from the observations.
The column density of gas-phase methanol is further boosted by the inclusion of grain-surface chemistry, reactive desorption, and fractal grains. 
However, even in the most optimistic case, the peak calculated column density at a time of 1~Myr remains around a factor of 2 to 3 lower than the disk-integrated value derived from the observations.  
The inclusion of vertical mixing in a disk model tailored to TW Hya may help to increase the column density further.

Models with grain-surface chemistry included predict an increase in gas-phase column density with radius, in contrast with that inferred from the observations.
We ran models in which we also scaled the initial ice-to-gas mass ratio by the dust-to-gas mass ratio, thereby beginning our calculations with the outer regions of the disk depleted in ice, and the inner disk midplane correspondingly enhanced in ice.  
There appears still sufficient dust-grain surface area in the outer disk to produce gas-phase methanol, with the column density in the outer disk gradually increasing until 10~Myr. 
Additional depletion of oxygen (and by extension CO) in the outer disk by a factor of 100, can reduce further the abundance and column density of methanol to below that constrained by the observations.

TW Hya remains the only disk around a Solar-type star within which we have successfully detected gas-phase methanol, and as such, it remains an important target for constraining the efficiency and efficacy of grain-surface chemistry in building chemical complexity in an environment similar to that present in the disk around the young Sun. The analysis conducted here provides compelling empirical evidence that the ice reservoir in TW Hya is compact, likely following the location of the large mm-sized dust grains, and confirms photodesorption, grain-surface chemistry and reactive desorption as important processes releasing methanol from the ice phase in cold regions of the disk.  However, despite the inclusion of successive amounts of chemical complexity, our static chemical disk model is unable to fully reproduce the magnitude and shape of the observed methanol column density profile.  We suggest that the inclusion of dynamic processes such as vertical mixing and the radial drift of dust are required to move sufficient amounts of methanol-rich ice from the colder regions of the disk into the luke-warm molecular layers in order to fully match the observed column densities. 

Our results demonstrate the importance of fully considering the interplay between complex chemical evolution in gas and ice, along with the growth and evolution of dust, in setting the abundance of prebiotic molecules.  The ability to fully understand the distribution and properties of these molecules in the comet- and planet-forming regions of protoplanetary disks offers a unique insight into the chemical complexity inherited by forming planets, and helps us understand how organic material can be delivered to young planetary systems.

\begin{acknowledgments}

We thank the anonymous referee for a thorough and constructive report.
J.D.I.~acknowledges support from an STFC Ernest Rutherford Fellowship (ST/W004119/1) and a University Academic Fellowship from the University of Leeds. 
C.W.~acknowledges financial support from the Science and Technology Facilities Council and UK Research and Innovation (grant numbers ST/X001016/1 and MR/T040726/1).
J.K.C. acknowledges support from the Kavli-Laukien
Origins of Life Fellowship at Harvard. 

This paper makes use of the following ALMA data: ADS/JAO.ALMA\#2016.1.00464.S ALMA is a partnership of ESO (representing its member states), NSF (USA) and NINS (Japan), together with NRC (Canada), MOST and ASIAA (Taiwan), and KASI (Republic of Korea), in cooperation with the Republic of Chile. The Joint ALMA Observatory is operated by ESO, AUI/NRAO and NAOJ. The National Radio Astronomy Observatory is a facility of the National Science Foundation operated under cooperative agreement by Associated Universities, Inc.

\end{acknowledgments}

\facility{ALMA}

\software{CASA \citep{CASA}, VISIBLE \citep{Loomis2018b}, Astropy \citep{astropy:2013, astropy:2018, astropy:2022}, Matplotlib \citep{Hunter:2007}, Numpy \citep{harris2020array}.}

\appendix

\section{Matched filter analysis}
\label{sec:filter}

\begin{figure*}
    \centering
    \includegraphics[width=\textwidth]{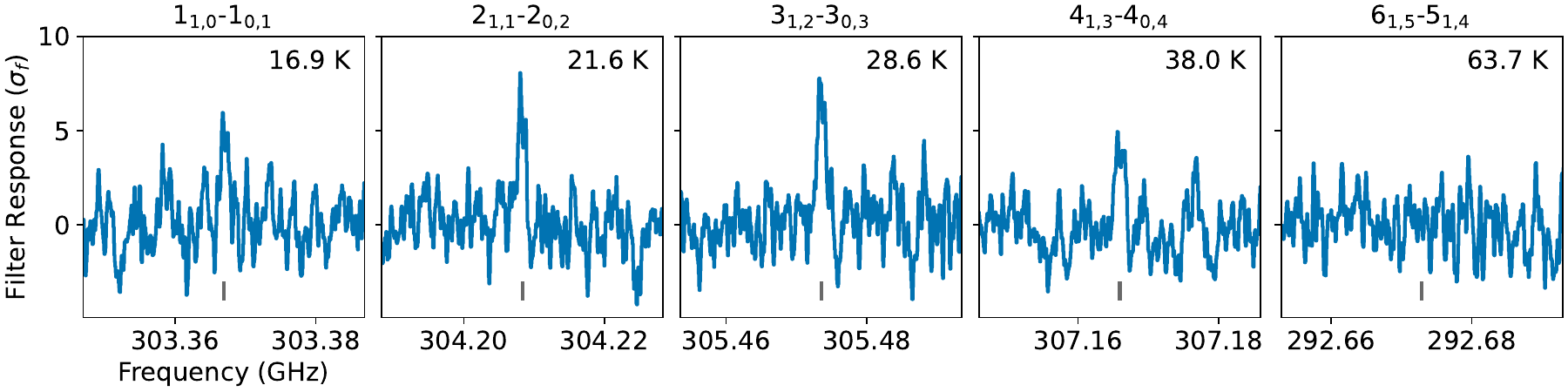}\\
    \vspace{0.3cm}
    \includegraphics[width=\textwidth]{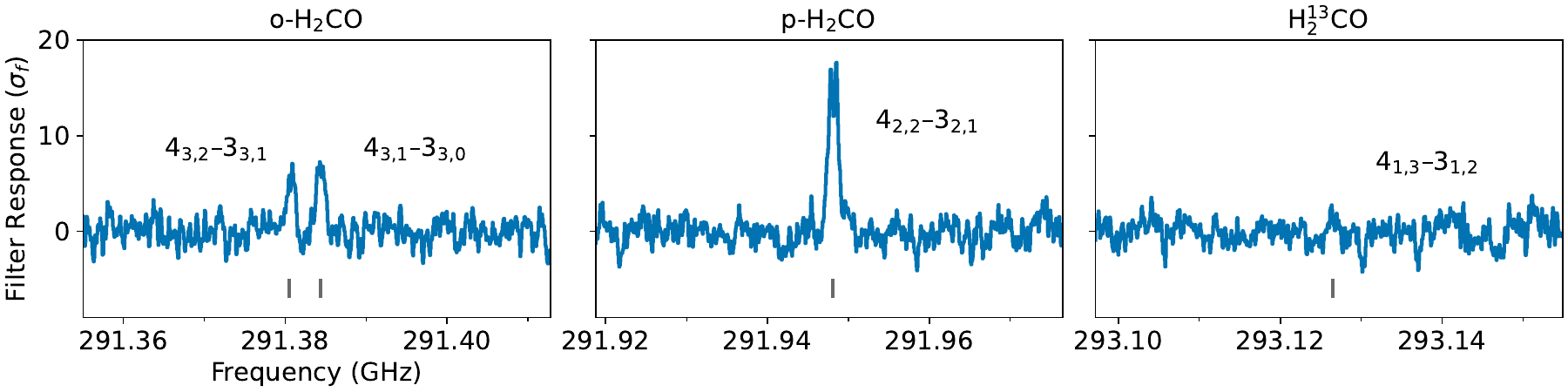}
    \caption{Matched filter response across the spectral windows containing methanol (\ce{CH3OH}) transitions using a Keplerian mask with a radius of 50~au (top) and the remaining spectral windows with a mask of radius 100~au (bottom). Rest frequencies of the labelled transitions are marked with vertical dashes.   See \citet{Jeroen2021} for an analysis of the \ce{H2CO} emission.}
    \label{fig:filter}
\end{figure*}

Given the high sensitivity of our observations, we performed the matched filter analysis across the entire spectral range of the data in order to confirm the detection of the methanol transitions, and search for serendipitous molecular line detections.  We note that due to a coarse velocity resolution (32\,km\,s$^{-1}$) it was not possible to perform this analysis on the continuum spectral window.  Figure \ref{fig:filter} shows the filter results for the targeted \ce{CH3OH} transitions and the additional spectral windows.  The latter shows detections of ortho- and para-\ce{H2CO} (see \citealt{Jeroen2021} for a full analysis) and a non-detection of H$_{2}^{13}$CO.  No other molecular transitions appear to be strongly detected across the other frequency ranges.  

\section{Channel Maps}
\label{sec:channelmaps}

\begin{figure*}
    \centering
    \includegraphics[width=0.769\textwidth, trim={0 1.95cm 0 2.1cm}, clip]{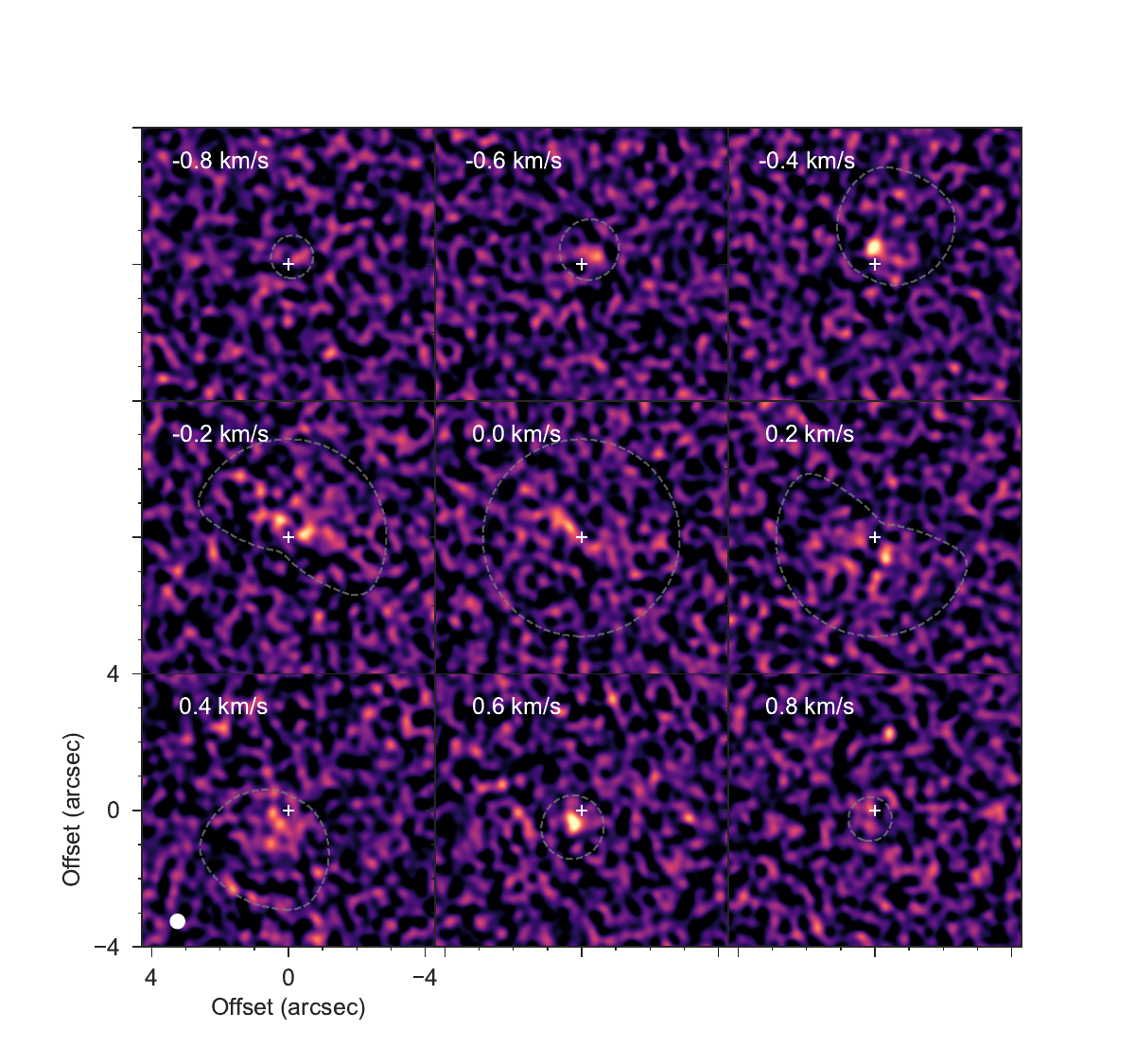}
    \includegraphics[width=0.769\textwidth, trim={0 0.6cm 0 2.1cm}, clip]{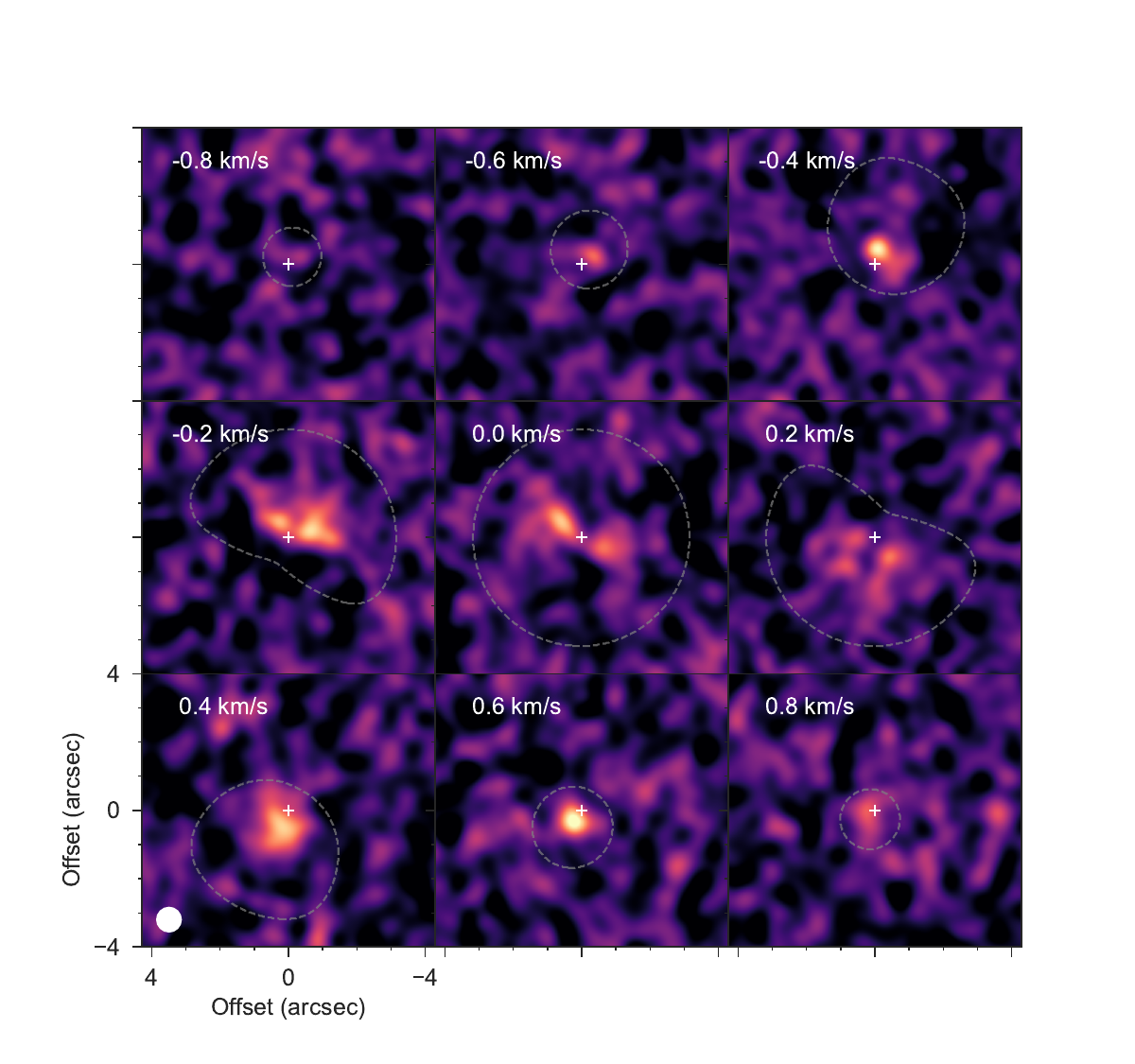}
    \caption{Stacked channel map of the methanol (\ce{CH3OH}) transitions with matched filter detections.  The top plot shows the 0\farcs4 observations, while the bottom plot shows the 0\farcs7 observations.  Each panel is shown on a linear stretch normalised to the maximum value across each image cube.}
    \label{fig:channel_map}
\end{figure*}

In Fig.~\ref{fig:channel_map} we show the channel maps for gas-phase methanol after stacking.  We stack only the four transitions with confirmed detections with the matched filter analysis (see Table~\ref{tab:molecular}).  
The top panels show the channel maps for the images with a 0\farcs4 beam, whereas the bottom panels show those after applying $uv$ tapering and smoothing to a beam of 0\farcs7.

\clearpage
\bibliography{TWHya}{}
\bibliographystyle{aasjournalv7}

\end{document}